\documentclass[sigconf]{acmart} 
\usepackage{tikz}
\usepackage{amsmath}

\usepackage{filecontents}
\usepackage{booktabs}
\usepackage{graphicx} 
\usepackage{caption}
\usepackage{multicol}
\usepackage{multirow}
\usepackage{enumitem}
\usepackage{array}
\usepackage{mathrsfs}
\usepackage{subcaption}
\usepackage[ruled,linesnumbered,boxed]{algorithm2e}
\usepackage{tabularx}
\usepackage{makecell}
\usepackage{threeparttable}
\usepackage{bm}
\usepackage{placeins}
\usepackage{xurl}
\usepackage{caption}
\usepackage{subcaption}
\usepackage{xcolor}
\newcommand{\rev}[1]{#1}

\allowdisplaybreaks[4]

\AtBeginDocument{%
  }

\copyrightyear{2026}
\acmYear{2026}
\setcopyright{cc}
\setcctype{by}
\acmConference[CCS '26] {Proceedings of the 2026 ACM SIGSAC Conference on Computer and Communications Security}{November 15--19, 2026}{The Hague, Netherlands.}
\acmBooktitle{Proceedings of the 2026 ACM SIGSAC Conference on Computer and Communications Security (CCS '26), November 15--19, 2026, The Hague, Netherlands}
\acmISBN{979-8-4007-2871-6/2026/11}
\acmDOI{10.1145/3830454.3846642}
\begin{document}


\title{Frequency-Domain Mixing Data Augmentation for Malicious Traffic Detection}




\author{Yuhao Yan}
\affiliation{%
  \institution{State Key Laboratory of Complex \& Critical Software Environment, Beihang University}
  \city{Beijing}
  \country{China}}
\email{yanyuhao@buaa.edu.cn}

\author{Bo Lang}
\authornote{Corresponding author.}
\affiliation{%
  \institution{State Key Laboratory of Complex \& Critical Software Environment, Beihang University}
  \city{Beijing}
  \country{China}}
\email{langbo@buaa.edu.cn}

\author{Xiangyu Li}
\affiliation{%
  \institution{State Key Laboratory of Complex \& Critical Software Environment, Beihang University}
  \city{Beijing}
  \country{China}}
\email{Li_xiangyu@buaa.edu.cn}








\begin{abstract} 
  The strong dynamics of network traffic often force malicious traffic detection models to handle out-of-distribution data. Typically, deep learning-based malicious traffic detection models require a large amount of high-quality training data. However, owing to challenges such as high labeling difficulty and resource consumption, existing datasets often suffer from insufficient diversity and fail to capture evolving traffic patterns, leading to poor out-of-distribution generalization ability of the trained models. Data augmentation has been widely adopted to improve data diversity and model generalization. Recently, frequency-domain mixing augmentation has shown promising performance because it effectively perturbs data while preserving key structural information. This approach shows potential for enhancing malicious traffic detection models. However, existing studies lack theoretical interpretation of the mixing mechanism, and do not adapt to the characteristics of network traffic. In this paper, we first conduct a theoretical analysis of the current frequency-domain mixing method, revealing its underlying principles and limitations. We further propose an improved frequency-domain mixing-based data augmentation method for network traffic data, which enhances the diversity of sequence features in network traffic and improves the out-of-distribution generalization of malicious traffic detection models. Extensive experiments on multiple artificial and real-world datasets demonstrate that our method substantially improves detection performance across diverse network environments and outperforms other data augmentation approaches.

\end{abstract}

\ccsdesc[500]{Security and privacy~Intrusion detection systems}


\keywords{Malicious Traffic Detection, Data Augmentation, Intrusion Detection} 


\maketitle

\section{Introduction} \label{sec:introduction}

Malicious network traffic detection is highly important for network security. Deep learning techniques are now widely used in malicious traffic detection \cite{HanCLZJL23,DialloP21,Fu0023,HanLLJLL24,HangLWX23,MirskyDES18,10.1145/3460120.3484585,290987} and have demonstrated excellent performance. 
Compared with traditional methods that rely on manually defined rules or statistical features for traffic classification \cite{AndersonM17,Barradas00SRM21}, deep learning-based approaches can automatically extract useful features from raw network traffic data.
In network communication, a session consists of a sequence of packets that carry important information. Consequently, many studies extract different feature sequences for traffic classification \cite{LinYHLX23,LinXGLSY22,LinXG21,10858569,LiuHXCL19,WuDQJ22}, such as packet length, inter-arrival time (IAT), and packet direction sequences. The sequences formed by these features contain rich information for identifying malicious traffic.

Network traffic inherently exhibits dynamics and distribution diversity \cite{areExisting,XieCDX00SZ23}, making it infeasible for training data to encompass all possible traffic types in the real world. These characteristics require that malicious traffic detection models possess strong out-of-distribution (OOD) generalization.
OOD can generally be categorized into \textit{covariate shift} and \textit{semantic shift} \cite{Generalized_OOD,areExisting}. In this work, we focus on \textit{covariate shift}, where the input distribution changes while the label distribution remains unchanged. 
Owing to the high cost of labeling and high resource consumption involved in collecting network traffic data \cite{GUERRA2022102810}, traffic in datasets is generally collected in artificially established, small-scale, and relatively stable environments \cite{SharafaldinLG18,9251211}, leading to discrepancies with real-world traffic and limited diversity. Models trained on such datasets tend to perform well on data following the same distribution as the training data (in-distribution data); however, when the data exhibit a distribution shift from the training data (OOD data), the model performance usually decreases sharply, indicating poor OOD generalization.

Data augmentation is an effective approach for increasing data diversity and improving OOD generalization \cite{GeirhosJMZBBW20}. As a form of time series data, network traffic feature sequences are amenable to time series augmentation techniques \cite{IglesiasTGMC23}. However, compared with generic time series data, network traffic exhibits more complex variation patterns \cite{XieCDX00SZ23}, which limits the effectiveness of conventional time series augmentation techniques.

Time series can be transformed into the frequency domain. In recent years, frequency-domain-based data augmentation has revealed distinct advantages in time series processing \cite{KimHK21,abs-2302-09292,ParkCZCZCL19,robusttad}. These methods identify trend characteristics and perturb data while preserving the key structural information. Inspired by Mixup \cite{ZhangCDL18}, frequency-domain mixing-based augmentation methods for time series data have been proposed. By mixing frequency-domain representations, they integrate structural information and periodic features to produce diverse samples and have achieved promising results in audio processing and time series forecasting \cite{KimHK21,abs-2302-09292}. Nevertheless, such methods lack theoretical analysis and do not consider the unique and complex properties of feature sequences in network traffic, leaving their application to malicious traffic detection a challenge.

Our goal is to enhance the OOD generalization of malicious detection models. We primarily target OOD traffic arising from two scenarios: variations of the same flows across different network environments \cite{XieCDX00SZ23}, and similar flows produced by identical or related services. Previous studies have shown that models focusing on low-frequency features tend to be more robust \cite{RahamanBADLHBC19,LiCRBBAPTP23,QianHYYLWGZ24}. Accordingly, we leverage frequency-domain mixing to encourage models to focus more on the low-frequency components of flows. We first conduct a theoretical analysis of existing frequency-domain mixing augmentation methods and show that they can be interpreted as matrix transformations in the time domain. Building on this foundation, we identify three problems when applying them to network traffic: (1) Original data with low diversity may fail to generate diverse samples; (2) directly using the frequency-domain mixing ratio to mix inter-class labels may be unsuitable; and (3) network traffic feature sequences are generally nonstationary, which makes the standard Fourier transform prone to distortion.
To address these issues, we propose an improved frequency-domain mixing-based augmentation method, which consists of two stages: pre-augmentation and core augmentation.  The pre-augmentation stage simulates network variations to preliminarily increase diversity. The core augmentation stage uses linear label mixing for inter-class samples and windowed Fourier transforms to handle nonstationary sequences. 


In summary, the main contributions of this paper are as follows:
\begin{enumerate}[left=0pt,label=$\bullet$]
    \item  We introduce frequency-domain mixing data augmentation for malicious traffic detection. We first provide a theoretical analysis of this kind of method to illustrate its  rationale, and then identify its limitations when applied to network traffic data.
    \item  Based on our theoretical analysis and the characteristics of network traffic, we propose a new frequency-domain mixing-based data augmentation method for feature sequences of network traffic, addressing these limitations through adding pre-augmentation, changing the label mixing pattern, and applying windowing to the Fourier transform.
    \item  We conduct extensive experiments on multiple artificial and real-world datasets to evaluate our method. The experimental results show that our proposed method effectively improves the OOD generalization of various mainstream models under different conditions and outperforms other data augmentation methods.
\end{enumerate}

\section{Related Work}\label{sec:related work}
This section reviews related work on malicious network traffic detection and data augmentation.
\subsection{Malicious Network Traffic Detection}
Early malicious traffic detection methods relied on handcrafted rules or statistical features \cite{AndersonM17,Barradas00SRM21,PanchenkoLPEZHW16}, which required extensive expert knowledge and struggled to detect unknown attacks. Deep learning-based detection techniques have been increasingly applied to malicious traffic detection, as they can automatically extract features from raw traffic data. Many studies utilize sequence information for traffic classification \cite{HanLLJLL24,LinYHLX23,LinXGLSY22,LinXG21,10858569,LiuHXCL19,WuDQJ22,CerasuoloNBACPR24,10.1145/3243734.3243768}. Liu et al. \cite{LiuHXCL19} proposed FS-Net for traffic classification, which is built based on a gated recurrent unit (GRU)  and can automatically extract features from packet length sequences. Shen et al. \cite{SAM} proposed SmartDetector, which utilizes packet length, IAT, and packet direction sequences to construct word embeddings and employs contrastive learning for malicious traffic detection. Although these methods have achieved promising results, they focus primarily on in-distribution (ID) data. In real-world network environments, packet sequences commonly exhibit variations such as reordering, duplication, and aggregation \cite{XieCDX00SZ23}, which leads to corresponding changes in the extracted feature sequences. As a result, real-world traffic often comprises OOD and unknown instances, which significantly degrade the performance of these models.

Some studies have focused on the detection of OOD malicious traffic. Han et al. \cite{HanLLJLL24} introduced a confidence branch into the classifier and classified traffic with confidence scores below a threshold as malicious, thereby improving the model’s ability to detect unknown attacks. Diallo et al. \cite{DialloP21} employed an adaptive clustering approach to enhance the robustness of the model. Unlike these approaches that design new models, our method improves OOD generalization through data augmentation.

\subsection{Data Augmentation} \label{sec:data augmentation}
Data augmentation improves model performance by generating new, semantically consistent samples from existing data and is widely used across various domains \cite{IglesiasTGMC23,abs-1904-11685,10.1145/3544558,YunHCOYC19}. Mixup \cite{ZhangCDL18} is a classic general augmentation technique. 
Its core idea is to generate new samples by performing linear interpolation between two existing samples and their corresponding labels.
Given two samples, $x_i$ and $x_j$, with their respective labels $y_i$ and $y_j$, Mixup generates a new sample $\tilde{x}$ and its label $\tilde{y}$, as follows:
\begin{align}
    \tilde{x} &= \lambda x_i + (1-\lambda) x_j, \label{eq:mixup_sample} \\
    \tilde{y} &= \lambda y_i + (1-\lambda) y_j, \label{eq:mixup_label}
\end{align}
where $\lambda$ is the mixing ratio.  
This operation helps to smooth the decision boundary and improve generalization. Another commonly used augmentation is random masking \cite{DevlinCLT19}, which masks certain features to prevent the model from overrelying on specific features.

For network traffic data, feature sequences can be treated as time series and are therefore suitable for time series augmentation techniques. Basic time-domain augmentations include scaling, flipping, jittering, and permutation \cite{IglesiasTGMC23}. For time series data, augmentation can be performed in the frequency domain, which can capture and manipulate global structural information and periodic features that are often difficult to obtain in the time domain. Park et al. \cite{ParkCZCZCL19} adopted time warping and frequency masking for audio data augmentation. Drawing on Mixup’s idea, both Kim et al. \cite{KimHK21} and Chen et al. \cite{abs-2302-09292} augmented sequence data by performing mixing in the frequency domain. This approach combines the frequency-domain features of different samples to generate new data. 
However, they did not provide a theoretical justification for the frequency-domain mixing augmentation. 
Our subsequent analysis (Section \ref{sec:limitations}) will reveal the limitations of this approach.

Data augmentation has also been applied in the field of network traffic classification.  Han et al. \cite{HanLLJLL24} augmented flow content features and pattern features by setting zero vectors and adding random perturbations, respectively. This approach can mitigate model overfitting, but provides limited improvement in OOD generalization. Xie et al. \cite{XieCDX00SZ23} proposed augmentation operations based on variations caused by network environments and used self-supervised learning to extract invariant features. 
However, their method is only applicable to packet length sequences, which limits its practicality.


\section{Problem Analysis} \label{sec:problem analysis}
In this section, we illustrate the problem of low diversity in network traffic datasets through an empirical analysis, investigate its underlying causes and adverse effects on the OOD generalization of malicious traffic detection models, and finally propose ideas to increase data diversity.

\subsection{Low Diversity in Network Traffic Datasets} \label{sec:case study}
\begin{figure}[htbp]
    \centering
    \includegraphics[width=0.8\linewidth]{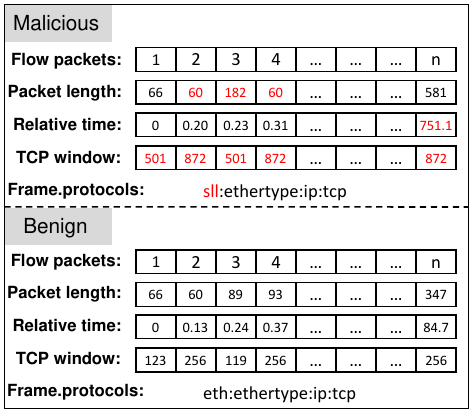}
    \vspace{-1mm}
    \caption{Examples of shortcuts we identified in the CICIDS-2017 and CIRA-CIC-DoHBrw2020 datasets. }
    \label{fig:bias_example}
    \vspace{-3mm}
\end{figure}

Datasets with low diversity are likely to contain shortcuts. We  investigate two widely used datasets, CIC-IDS2017 \cite{SharafaldinLG18} and CIRA-CIC-DoHBrw2020 (CIC-DoH) \cite{9251211}, and identify several shortcuts that models may exploit. We refer to these features as ``shortcuts'' because, although they can help models achieve strong performance quickly, they are in fact fragile and non-robust features. Some of the identified shortcuts are illustrated in Figure \ref{fig:bias_example}.

In the botnet traffic of CIC-IDS2017, the packet length sequence ``[60, 182, 60]'' serves as a conspicuous pattern. Simply detecting this pattern via string matching can already achieve a recall of 88.57\% and a precision of 93.34\%. In CIC-DoH:
\begin{enumerate}[left=0pt,label=$\bullet$]
    \item Malicious flows tend to have much longer durations, meaning that the “relative time  from the first packet” of the last packet is typically large (over 500 seconds), whereas most benign flows have significantly shorter durations.
    \item The TCP window size often takes on a small set of fixed values; for example, in malicious traffic generated by dnscat2, this value is 501, which is absent in most benign flows.
    \item Parsing the “frame.protocols” field via Wireshark reveals that benign traffic uses “eth” at the link layer, whereas malicious traffic uses “sll”.
\end{enumerate}
These features are inherent biases introduced by limitations in the data collection process and are not applicable in real-world scenarios. In practice, a natural packet loss event can easily disrupt the sequence ``[60, 182, 60]''; malicious flows do not necessarily exhibit such long durations; TCP window sizes usually fluctuate frequently; the ``sll'' in link layer is from the use of the Linux Cooked Capture pseudo-protocol and is entirely independent of the actual properties of malicious traffic.

The presence of these shortcuts indicates that the diversity of these datasets is insufficient, such that these inherently fragile features lack diverse variation patterns and instead become key cues for distinguishing malicious traffic. Models trained in this way tend to suffer significant performance degradation in real-world scenarios. In fact, the LUCID model proposed by Doriguzzi-Corin et al. \cite{LUCID} utilizes features including relative time, TCP window size, and ``frame.protocols''. Our experiments on dataset $\mathcal{D}_1$ (detailed in Section \ref{sec:experiments}) show that LUCID achieves a high F1 score of 94.89\% and a low false positive rate (FPR) of 0.53\% by leveraging these features; however, simply perturbing or removing these features causes the F1 score to drop below 50\% and the FPR to rise above 1\%.

In addition to shortcuts, these datasets also exhibit high traffic similarity. For instance, an LSTM trained on only 1\% of the CIC-DoH dataset already achieves an F1-score of approximately 93\% and an FPR of 1\%. We believe such low diversity is a common problem in manually collected network traffic datasets. Beyond the improper experimental setup, we identify two main reasons for the lack of diversity: 
\textbf{(1) Inherent network protocol behavior}. Many protocols exhibit repetitive patterns. For example, TCP acknowledgment mechanism generates numerous ACK packets, which increases the similarity between different sequences.
\textbf{(2) Large amounts of traffic from the same application or similar services}. Traffic generated by the same application or similar services is likely to be highly similar. For example, when a user makes multiple visits to certain web pages, the traffic may have only minor differences in the request body.

These two factors are inherent to network traffic. Due to difficulties in labeling and high resource costs, traffic is generally captured in artificially controlled small-scale environments. This setting may significantly exacerbate the second situation mentioned above, resulting in low diversity of the obtained data. When data diversity is low, deep learning models tend to exploit shortcuts in the data \cite{GeirhosJMZBBW20}. As evident from our experiments in Sections \ref{sec:augment performance} and \ref{sec:impact on in-distribution}, models trained on such datasets often exhibit high ID performance but poor OOD generalization, which severely limits their practical utility.

\subsection{Thoughts to Solve the Problem} \label{sec:ideas to solve}
To enhance the model's OOD generalization, more robust features are required. Some studies have shown that focusing on low-frequency features can make models more robust \cite{RahamanBADLHBC19,LiCRBBAPTP23,QianHYYLWGZ24}. In the time domain, low-frequency components typically represent the overall trajectory and trend of a sequence. 

We primarily target OOD traffic (represented as packet feature sequences) arising from two scenarios: variations of the same flows across different network environments, and similar flows produced by identical or related services. For example, the former may include packet duplications due to retransmissions or out-of-order arrivals caused by the environment \cite{XieCDX00SZ23} (although TCP protocol can handle issues such as duplication to ensure the consistency of application-layer data, capture usually occurs before the packets enter the network protocol stack), while the latter may include malware and its variants, or applications with similar behaviors. In both cases, the resulting traffic mainly exhibits local variations (e.g., changes and additions) that have minimal effect on the overall trend of the sequence; in other words, their low-frequency components remain relatively similar.

\begin{figure}[htbp]
    \centering
    \includegraphics[width=0.8\linewidth]{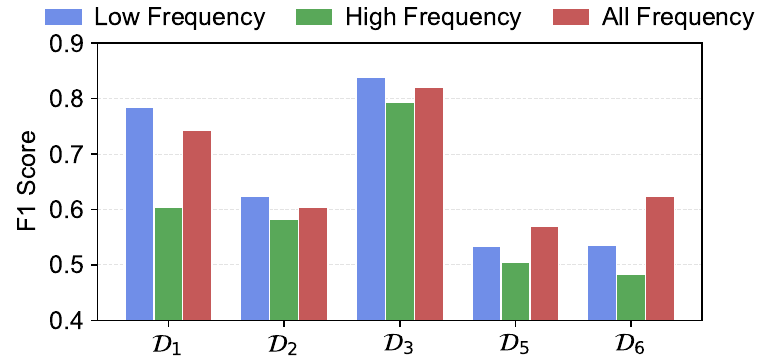}
    \vspace{-2mm}
    \caption{F1 scores of CNNs using low-, high-, and all-frequency features for malicious traffic detection.}
    \label{fig:low_frequency_robust}
    \vspace{-3mm}
\end{figure}

\rev{We can obtain frequency-domain representation using the fast Fourier transform (FFT) and transform it back to the time domain using the inverse fast Fourier transform (IFFT). A sequence can be decomposed into low-frequency components capturing smooth trends, and high-frequency components reflecting volatile changes. We use all datasets described in Section \ref{sec:experimental setup} except $\mathcal{D}_4$ (because its small number of malicious traffic samples may introduce substantial bias) to conduct a preliminary experiment to validate the robustness of low-frequency features. For each sample, we generate low- and high-frequency versions by retaining only the first 50\% and the last 50\% of its frequency-domain components, respectively, and transforming them back to the time domain. The resulting samples are then used to train two separate CNNs for OOD testing. For comparison, we also train a CNN using all frequency components. The experimental results are shown in Fig. \ref{fig:low_frequency_robust}. Models using only low-frequency features perform comparably to (and sometimes exceed) those using all-frequency features, and consistently outperform those using only high-frequency features. This suggests that low-frequency information is more robust, whereas high-frequency information may sometimes act as interference to some extent. }

To force the model to focus on low-frequency features, one idea is to concatenate the low-frequency components of one flow with the high-frequency components of another and then transform it back to the time domain via inverse fast Fourier transform (IFFT). This combination preserves the original flow’s low-frequency features while introducing variations through the high-frequency components, thereby perturbing the sequence. Such perturbations may disrupt other non-robust features (especially shortcuts) while keeping the low-frequency features relatively stable. In this way, the model is encouraged to learn the low-frequency characteristics of the data, thus enhancing its robustness. \rev{The above approach is an existing frequency-domain mixing data augmentation \cite{KimHK21,abs-2302-09292}. However, we find that this method is not directly suitable for network traffic. We will analyze this method and propose improvements based on it in the next section.}

\rev{We obtain low-frequency features through the FFT and treat them as robust features. Based on these features, we perturb other components to generate augmented samples. Therefore, obtaining robust features is an important part of our method. Currently, feature extraction and representation learning can also be used to obtain data features. However, feature extraction typically relies on manually designed features, while representation learning obtains latent features through model training, with its quality depending on task design and model capacity. Moreover, it is often difficult to identify robust feature components from the features produced by either of these two approaches. Unlike feature extraction or representation learning, our method obtains robust features directly via FFT, without relying on any model or task. Although the assertion of the robustness of low-frequency features is not our contribution, our novelty lies in recognizing the relative stability of overall trends in network traffic, introducing frequency-domain analysis to represent these stable trends as robust low-frequency features, and leveraging this property to design a new input augmentation mechanism for malicious traffic detection.}

\section{Method} \label{sec:Method}
In this section, we first formally define the existing frequency-domain mixing data augmentation method and conduct a theoretical analysis. We further reveal the limitations of this method. Finally, we address these limitations and propose our improved frequency-domain mixing-based augmentation method.

\subsection{Theoretical Analysis of the Frequency-Domain Mixing Augmentation Method}
\subsubsection{Formal Definition} \label{sec:Formal Definition}
Based on the ideas proposed in Section \ref{sec:ideas to solve}, we first introduce the conventional frequency-domain mixing method in a formalized form.

Let $x_1$ and $x_2$ be two different time-domain samples (sequences), each of length $N$, with corresponding labels $y_1$ and $y_2$. We can obtain their frequency-domain representations, $X_1$ and $X_2$, by performing the FFT $\mathcal{F}$. Since the sequence generated by the FFT has the same length as the original sequence, $X_1$ and $X_2$ also have a length of $N$. For a sequence $X$, we use $X[a]$ to denote the element at index $a$, and $X[a:b]$ to denote the subsequence from index $a$ up to (but not including) index $b$. The index is 0-based, so a sequence of length $N$ has indices from $0$ to $N-1$. For example, if $X=[0,1,2,3]$, then $X[0]=0$ and $X[1:3]=[1,2]$.

We now concatenate their frequency-domain components. Considering that the sequences we process are real-valued, their Fourier transforms possess conjugate symmetry. Therefore, we consider only the first half of the elements in the frequency domain, i.e., those from index $0$ to $\lfloor \frac{N}{2} \rfloor$ (where $\lfloor \cdot \rfloor$ denotes the floor function). We randomly select a frequency-domain splitting ratio $\lambda$ where $0 < \lambda < 1$, which yields a splitting point $l = \lfloor \lfloor \frac{N}{2} \rfloor \lambda \rfloor$. Accordingly, the frequency-domain representation of the new sample $X$ is as follows:
\begin{align}
X[0:l] &= X_1[0:l], \label{eq:freqmix begin} \\
X[l:\lfloor \frac{N}{2} \rfloor + 1] &= X_2[l:\lfloor \frac{N}{2} \rfloor + 1].
\end{align}
The remaining part of $X$ is obtained through conjugate symmetry, ensuring that the reconstructed sequence remains a real-valued sequence. For the $k$-th element, we have:
\begin{equation}
X[k] = X^*[N-k], \  \lfloor \frac{N}{2} \rfloor + 1 \le k < N, \label{eq:conjugate stmmetry}
\end{equation}
where  $^*$ denotes the complex conjugate. Finally, we apply the IFFT to obtain the new sample in the time domain:
\begin{equation}
    x=\mathcal{F}^{-1} (X).     \label{eq:ifft}
\end{equation}
The label $y$ of the new sample is also obtained by mixing the labels of the original two samples in the same proportion:
\begin{equation}
    y=\lambda y_1+(1-\lambda) y_2. \label{eq:freqmix end}
\end{equation}

Notably, Kim et al. \cite{KimHK21} and Chen et al. \cite{abs-2302-09292} applied this frequency-domain mixing in different scenarios, resulting in minor differences between their operations and the operation we present here. However, the core of their methods remains the same, as described by Eqs. \eqref{eq:freqmix begin}--\eqref{eq:freqmix end}. Nevertheless, Kim et al. \cite{KimHK21} and Chen et al. \cite{abs-2302-09292} described only the algorithm and did not provide a theoretical basis for this approach. We will now analyze this approach theoretically.

\subsubsection{Theoretical Analysis} \label{sec:Theoretical Analysis} 
Based on Eqs. \eqref{eq:freqmix begin}--\eqref{eq:freqmix end} in Section \ref{sec:Formal Definition}, we set the splitting ratio $\lambda=2\alpha$; thus, $l=\lfloor 2\alpha \lfloor \frac{N}{2} \rfloor \rfloor$. For convenience, we assume that both $\lfloor \frac{N}{2} \rfloor$ and $\alpha N$ are integers; thus, $l=\alpha N$. We define a sequence $H_1$ of length $N$ in the frequency domain, whose $k$-th elements are defined as follows:
\begin{equation}
H_1[k] =
\begin{cases}
1, & 0 \le k < \alpha N \\
0, & \alpha N \le k \le (1-\alpha) N \\
1, & (1-\alpha) N < k < N
\end{cases}, \label{eq:h1}
\end{equation}
and set $H_2=E'-H_1$, where $E'=[1,1,1,…,1]$ is the all-ones sequence of length $N$.
$H_1$ and $H_2$ act as masks, replacing our previous ``truncation-concatenation'' operation. Let  $\cdot$ denote the Hadamard product (i.e., elementwise product); then, the frequency-domain mixing operation can be expressed as follows:
\begin{align}
X &= X_1 \cdot H_1 + X_2 \cdot H_2 \notag \\
  &= X_1 \cdot H_1 + X_2 \cdot (E' - H_1) \notag \\
  &= X_1 \cdot H_1 + X_2 - X_2 \cdot H_1. \label{eq:mix_operation}
\end{align}
We subsequently apply the inverse Fourier transform to $X$. Since the inverse Fourier transform is linear, we have the following:
\begin{align}
x &= \mathcal{F}^{-1}(X) \notag \\
  &= \mathcal{F}^{-1}\big(X_1 \cdot H_1 + X_2 - X_2 \cdot H_1 \big) \notag \\
  &= \mathcal{F}^{-1}(X_1 \cdot H_1) + \mathcal{F}^{-1}(X_2) - \mathcal{F}^{-1}(X_2 \cdot H_1). \label{eq:mix_operation_ifft}
\end{align}
According to the convolution theorem, the product in the frequency domain is equivalent to the convolution in the time domain. For finite sequences, the convolution here is cyclic rather than linear. Let the inverse Fourier transforms of $H_1$ and $H_2$ be $h_1$ and $h_2$, respectively. Then, the above expression can be transformed into the following form:
\begin{align}
x &= x_1 \ast h_1 + x_2 - x_2 \ast h_1 \notag \\
  &= x_2 + (x_1 - x_2) \ast h_1, \label{eq:x_conv_h1}
\end{align}
where $\ast$ denotes cyclic convolution.
This equation indicates that the new sample $x$ is actually obtained by superimposing a ``perturbation'' onto the original sample, which is the convolution of the difference between the two samples and $h_1$. Notably, $h_1$ here is a deterministic sequence obtained by applying the inverse Fourier transform to $H_1$.

The form of Eq. \eqref{eq:x_conv_h1} is notably similar to the augmentation strategy of Mixup \cite{ZhangCDL18}. A comparison of Eq. \eqref{eq:mixup_sample} and Eq. \eqref{eq:x_conv_h1} reveals that the augmentation operation of Mixup is actually a special case of frequency-domain mixing augmentation. If we set $h_1=[\lambda,0,0,…,0]$ in Eq. \eqref{eq:x_conv_h1}, we can obtain exactly the same result as Mixup. In essence, the frequency-domain mixing method generates new samples via convolution.

In Eq. \eqref{eq:x_conv_h1}, the cyclic convolution operation can be replaced by multiplication with a circulant matrix. If $h_1=[c_0,c_1,…,c_{N-1}]$, its corresponding circulant matrix $A$ is defined as follows:
\begin{equation}
A =
\begin{bmatrix}
c_0      & c_{N-1} & \cdots & c_2      & c_1 \\
c_1      & c_0     & c_{N-1}& \cdots   & c_2 \\
\vdots   & c_1     & c_0    & \ddots   & \vdots \\
c_{N-2}  & \vdots  & \ddots & \ddots   & c_{N-1} \\
c_{N-1}  & c_{N-2} & \cdots & c_1      & c_0
\end{bmatrix}.
\label{eq:circulant}
\end{equation}
For convenience, we represent $x,x_1,x_2$ as column vectors. Eq. \eqref{eq:x_conv_h1} can then be rewritten as follows:
\begin{equation}
    x=Ax_1+(E-A) x_2.
    \label{matrix form}
\end{equation}
Here, $E$ is the identity matrix. Assuming that the distribution of the original data is $\mathcal{D}$, $x$ can be considered a transformed sum of two instances drawn from $\mathcal{D}$. 

Eq. \eqref{matrix form} is the theoretical formulation of the frequency-domain mixing method, showing that frequency-domain mixing performs mixing in the time domain via matrix transformation. The matrix transformation increases the diversity of generated samples and expands their coverage in the feature space, thereby improving generalization. By comparison with Mixup again, we see that Eq. \eqref{matrix form} actually replaces the scalar multiplication in Mixup with matrix multiplication. If $A$ is set as a diagonal matrix with all diagonal elements equal to $\lambda$, Eq. \eqref{matrix form} becomes equivalent to Mixup. 
\begin{figure}[htbp]
    \centering
    \begin{subfigure}[b]{0.49\linewidth}
        \centering
        \includegraphics[width=\linewidth]{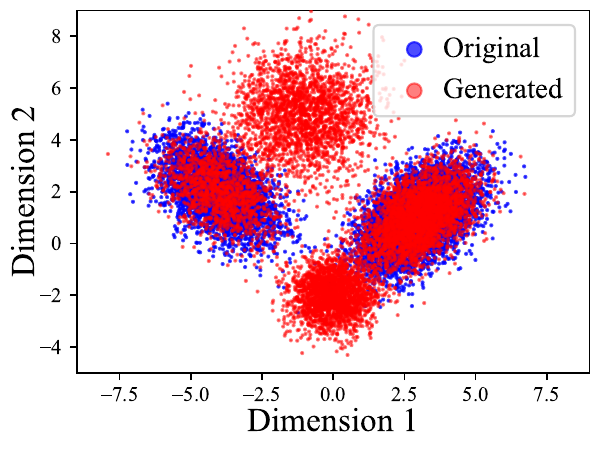}
        \caption{Frequency-domain mixing.}
        \label{fig:gaussian_distribution_a}
    \end{subfigure}
    \hfill
    \begin{subfigure}[b]{0.49\linewidth}
        \centering
        \includegraphics[width=\linewidth]{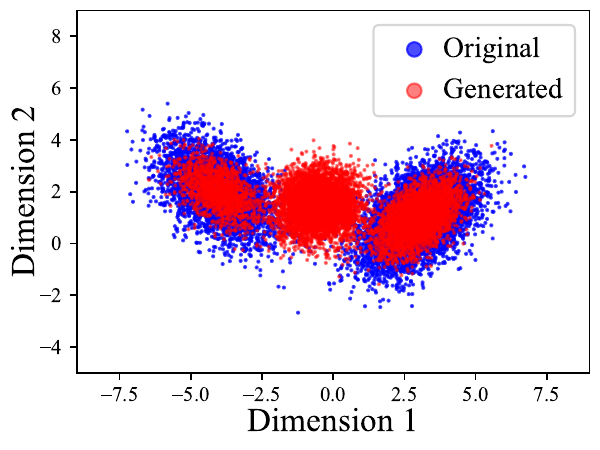}
        \caption{Mixup.}
        \label{fig:gaussian_distribution_b}
    \end{subfigure}
    \vspace{-2mm}
    \caption{Sample distributions generated by frequency-domain mixing and Mixup.}
    \label{fig:gaussian_distribution}
\end{figure}
To better illustrate that the matrix transformation in Eq. \eqref{matrix form} expands the data distribution, we provide an example with visualization, in which a two-dimensional Gaussian mixture distribution with two components is used as the original distribution, and two data augmentations, frequency-domain mixing and Mixup, are applied respectively to generate new samples. In Figure \ref{fig:gaussian_distribution}, blue points denote original samples and red points denote augmented samples. As shown in Figure \ref{fig:gaussian_distribution_a}, frequency-domain mixing produces samples with lower density in the high-concentration regions of the original data, while also appearing in areas previously uncovered by the original distribution. We compare frequency-domain mixing with Mixup, as shown in Figure \ref{fig:gaussian_distribution_b}. Mixup generates new samples primarily through linear interpolation, resulting in samples that are not only located within the originally covered regions but also concentrated along the line segments between the two Gaussian components. In contrast, the frequency-domain mixing approach applies a matrix transformation, enabling the new samples to spread across a much broader space.


\subsection{Limitations of the Existing Frequency-Domain Mixing Method} \label{sec:limitations}
Based on the above analysis and the unique characteristics of network traffic, the existing frequency-domain mixing method has the following limitations when applied to malicious traffic detection.

\begin{figure*}[hbtp]
    \centering
    \includegraphics[width=0.94\linewidth]{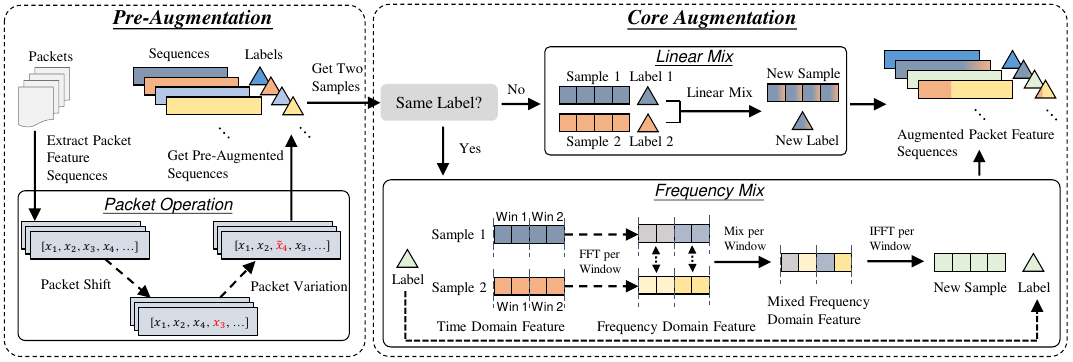}
    \vspace{-2mm}
    \caption{The proposed improved frequency-domain mixing-based method.}
    \label{fig:framework}
    \vspace{-3mm}
\end{figure*}

\noindent \textbf{(1) Diminished Effectiveness on Data with Originally Low Diversity.} 
Frequency-domain mixing does not guarantee increased data diversity. If the original data lack diversity and are overly concentrated, the generated samples will likely remain within the regions where the original data are densely distributed. In such cases, the generated samples will be highly similar to existing samples and thus fail to enrich the dataset. 
For instance, for a multivariate Gaussian distribution, the transformation in Eq. \eqref{matrix form} does not alter the mean vector, meaning that the concentration region of the new samples is identical to that of the original data. Consequently, the diversity of the data does not increase. Furthermore, depending on the properties of matrix $A$, the covariance of the distribution may decrease in some cases. This implies that the newly generated data become more concentrated, with even worse diversity than that of the original data. Such augmentation provides less benefit to the model. Section \ref{sec:case study} shows that network traffic datasets generally suffer from poor diversity, which limits the effectiveness of the frequency-domain mixing method to some extent.

\noindent \textbf{(2) Inappropriate Inter-Class Label Mixing.}
With respect to mixing-based data augmentation methods such as Mixup, an inappropriate mixing ratio can negatively impact model performance \cite{DBLP:journals/tnn/MaiHCSS22}. Mixup linearly mixes two samples (Eqs. \eqref{eq:mixup_sample}--\eqref{eq:mixup_label}) and simultaneously mixes their labels according to the same ratio. This intuitive mixing strategy has been proven effective. In the existing frequency-domain mixing method, the label mixing ratio is determined by the proportion of the segment lengths taken from each sample in the frequency domain (Eqs. \eqref{eq:freqmix begin}–\eqref{eq:freqmix end}). While this process is intuitive in the frequency domain, once the mixed data are transformed back to the time domain, models trained on time-domain data may struggle to capture this relationship. Therefore, this approach may not be appropriate. Considering Eq. \eqref{matrix form}, if matrix $A$ is a diagonal matrix where all the diagonal elements are $\lambda$, this approach is equivalent to Mixup with a mixing ratio $\lambda$. However, in frequency-domain mixing, matrix $A$ is a circulant matrix derived from the inverse Fourier transform of the sequence $H_1$ as defined in Eq. \eqref{eq:h1}, which does not have a simple diagonal matrix form. Therefore, directly applying a mixing ratio $\lambda$ for inter-class label mixing may have adverse effects.

\noindent \textbf{(3) Negative Impact of Nonstationary Sequences.}
A nonstationary sequence has frequency components that change over time. In network communications, even within a single session, the local behavior patterns may change according to the specific content, making most network traffic data nonstationary sequences. Applying the Fourier transform to such sequences results in the loss of temporal information and fails to capture the pattern of how frequency components evolve over time. In such cases, frequency-domain mixing may lead to deviations between the retained trend information and the real trend information, thereby degrading model performance.

\subsection{Improved Frequency-Domain Mixing-Based Method} \label{sec:our method}
\subsubsection{Framework}
Based on the analysis in Section \ref{sec:limitations}, we propose an improved frequency-domain mixing-based method. The following measures are adopted to address the three problems mentioned in Section \ref{sec:limitations}:
(1) Given that the poor diversity of the original data leads to generated samples being similar to existing samples, we perform pre-augmentation before implementing the core frequency-domain mixing method to preliminarily increase data diversity.
(2) To address the problem of an inappropriate mixing ratio for inter-class labels, we avoid frequency-domain mixing for such samples because its transformation matrix is not diagonal, making the appropriate label mixing ratio unclear. Mixup’s linear mixing has been widely validated, so we adopt it for inter-class samples. Since intra-class labels remain unchanged regardless of the mixing ratio, frequency-domain mixing is still applied in this case.
(3) To address the impact caused by nonstationary sequences, we apply windowing to the Fourier transform, which first windows the sequences and then executes the Fourier transform and conducts frequency-domain mixing augmentation independently within each window.


The overall framework of the proposed method is illustrated in Figure \ref{fig:framework}. Our method consists of two stages: pre-augmentation and core augmentation. The pre-augmentation stage first extracts packet feature sequences, then applies packet shift and packet variation operations to simulate realistic network changes. The subsequent core augmentation selects two samples each time: if their labels differ, the two samples and their labels are linearly mixed;  otherwise, the sequences are divided into windows, and frequency-domain mixing is applied within each window, with the original label preserved. After core augmentation is completed, the final augmented sequences are obtained.

\subsubsection{Pre-Augmentation}

Prior to the core augmentation stage, we first perform pre-augmentation on the original data. The primary purpose of pre-augmentation is to simulate the variations in real-world network traffic, thereby increasing the diversity of the original data and broadening its distribution to enhance the effectiveness of core augmentation. Generally, elements in a feature sequence are derived from individual packets, meaning that changes in packets lead to corresponding variations in the resulting features. Therefore, we design pre-augmentation approaches based on packet variation patterns to ensure that our method is applicable to a wide range of sequential features. Considering the potential variations in network packets under real-world network conditions \cite{XieCDX00SZ23}, we employ two operations for pre-augmentation: \textit{packet shift} and \textit{packet variation}. \textit{Packet shift} randomly shifts the packets at selected positions by certain offsets. \textit{Packet variation} randomly perturbs certain attributes of packets at the selected positions.
We randomly select several positions within a sequence to perform these operations. We constrain these two augmentation operations to small magnitudes. Excessive changes may compromise label consistency between the original and augmented sequences, thereby hindering the effectiveness of the subsequent core augmentation stage.


    

\subsubsection{Core Augmentation}
After the pre-augmented data are obtained, we select two samples each to perform our core augmentation operations. If the two samples have different labels, we directly apply linear mixing to the two samples and their labels (Eqs. \eqref{eq:mixup_sample}--\eqref{eq:mixup_label}). If the two samples share the same label, our windowed frequency-domain mixing method is executed. We first use a windowed Fourier transform to extract the frequency-domain information from the sequences. Given a window length $w$, the Fourier transform is applied to each subsequence of $w$ elements to acquire the frequency-domain features. For each pair of corresponding windows from the two samples, frequency-domain mixing defined by Eqs. \eqref{eq:freqmix begin}--\eqref{eq:conjugate stmmetry} is performed. An inverse Fourier transform (Eq. \eqref{eq:ifft}) is subsequently applied to each window to transform the result back to the time domain, yielding the final new sequence. 
The entire data augmentation method is shown in Algorithm \ref{algorithm1}.
\begin{algorithm}[b]
\DontPrintSemicolon
  \SetAlgoNoLine
  \caption{Our Data Augmentation Method}
  \label{algorithm1}
  \KwIn{Packet feature sequences $x_1,x_2$ (padded or truncated to same length) and their labels $y_1,y_2$, window length $w$, mixing ratio $\lambda$, the ratio of elements to be pre-augmented $r$}
  \KwOut{New generated sequence $x$ and its label $y$}
  \tcp{Pre-augmentation}
Packet\_Shift($x_1, x_2, r$)\;
Packet\_Variation($x_1, x_2, r$)\;

\tcp{Core Augmentation}
\If{$y_1 = y_2$}{
    $i \gets 0$\;
    $y \gets y_1$\;
    $X_1[0 \dots length(x_1)] \gets 0$, $X_2[0 \dots length(x_2)] \gets 0$\; 
    $X[0 \dots length(x_1)] \gets 0$, $x[0 \dots length(x_1)] \gets 0$\; 
    
    \While{$i < \text{length}(x_1)$}{
        $X_1[i \dots i+w] \gets \text{FFT}(x_1[i \dots i+w])$\;
        $X_2[i \dots i+w] \gets \text{FFT}(x_2[i \dots i+w])$\;
        \tcp{Mix frequency according to Eqs. \eqref{eq:freqmix begin}-\eqref{eq:conjugate stmmetry}}
        $X[i \dots i+w] \gets \text{Mix\_Frequency}(X_1[i \dots i+w], X_2[i \dots i+w], \lambda)$\;
        $x[i \dots i+w] \gets \text{IFFT}(X[i \dots i+w])$\;
        $i \gets i + w$\;
    }
}
\Else{
    $x \gets \lambda x_1 + (1-\lambda)x_2$\;
    $y \gets \lambda y_1 + (1-\lambda)y_2$\;
}
\Return{$x, y$}\;
\end{algorithm}

\subsubsection{Semantic Preservation of Our Method}
\rev{Data augmentation generally should not alter the semantics of data. Here, we interpret semantic invariance as the invariance of task-relevant semantics, i.e., the invariance of the true label distribution conditioned on the input sample. In our method, pre-augmentation simulates certain variations in real network traffic and constrains these variations to a small magnitude, thus generally preserving the original semantics; the inter-class linear mixing strategy is the same as Mixup, whose effectiveness has been validated by previous studies \cite{ZhangDK022,mixuptraining}. Therefore, we focus on analyzing the semantic preservation of the intra-class frequency-domain mixing strategy.}

\rev{In the frequency domain, given a cutoff frequency index $l$, components with indices smaller than $l$ are regarded as low-frequency components, while the remaining components are regarded as high-frequency components.  In the time domain, a sequence $x$ can be decomposed as $x=x^L+x^H$, where $x^L$ is reconstructed from the low-frequency components of $x$, and $x^H$ from the remaining high-frequency components. This decomposition is unique, i.e., there is a one-to-one correspondence between $x$ and the pair $(x^L,x^H)$. Since the Fourier basis functions are mutually orthogonal, modifying the high-frequency components does not affect the low-frequency components and their corresponding time-domain reconstruction $x^L$, and vice versa. Consequently, when we concatenate the low-frequency components of $x_1$ with the high-frequency components of $x_2$, the resulting new sample $\tilde{x}$ in the time domain can be described as:
\begin{equation}
    \tilde{x}=\tilde{x}^L+\tilde{x}^H=x^L_1+x^H_2,\quad \text{where }\tilde{x}^L=x_1^L,\ \tilde{x}^H=x_2^H.   \label{eq:time domain decompose}
\end{equation}}
\rev{Let $p(y|x)$ denote the true label distribution of sample $x$. Since $x$ can be uniquely decomposed into the pair $(x^L,x^H)$, $p(y|x)$ can equivalently be written as $p(y|x^L,x^H)$. We use the total variation (TV) distance $TV(\cdot,\cdot)$ to measure the discrepancy between two distributions, which is commonly used in machine learning theory \cite{abs-2004-11829-TV}. We assume that low-frequency features are more robust, meaning that, for most samples, the label distribution can largely be inferred from low-frequency features and is relatively insensitive to variations in the high-frequency components, so that perturbing the high-frequency components does not substantially alter the label distribution. For a given class label $c\in\{0,1\}$, let $x$ and $x'$ be independently drawn from the same class-conditional distribution $p(x|y=c)$. This assumption can be formally stated as: there exists a sufficiently small value $\epsilon$ such that
\begin{equation}
    \mathbb{E}_{x,x'\sim p(x|y=c)} [TV(p(y|x^L,x^H),p(y|x^L,x'^H))]<\epsilon,  \label{eq:TV robust}
\end{equation}
where $\mathbb{E}$ denotes the expectation.} 

\rev{The augmented sample $\tilde{x}$ is generated from two samples $x_1$ and $x_2$ independently drawn from $p(x|y=c)$. According to Eq. (\ref{eq:time domain decompose}), the low- and high-frequency sequences in the time-domain decomposition of $\tilde{x}$ are $x_1^L$ and $x_2^H$, respectively. Combining this with Eq. (\ref{eq:TV robust}), the expected discrepancy in the label distribution between $\tilde{x}$ and $x_1$ is given by:
\begin{align}
    &\mathbb{E}_{x_1,x_2\sim p(x|y=c)}[TV(p(y|x_1),p(y|\tilde{x}))] \notag \\
    =\ &\mathbb{E}_{x_1,x_2\sim p(x|y=c)}[TV(p(y|x_1^L,x_1^H),p(y|\tilde{x}^L,\tilde{x}^H))] \notag  \\
    =\ &\mathbb{E}_{x_1,x_2\sim p(x|y=c)}[TV(p(y|x_1^L,x_1^H),p(y|x_1^L,x_2^H))] < \epsilon. \label{eq:semantic}
\end{align}
Eq. (\ref{eq:semantic}) shows that the expected TV distance between the label distribution $p(y|\tilde{x})$ of our augmented sample $\tilde{x}$ and that of the original sample $x_1$, $p(y|x_1)$, is also no more than $\epsilon$. This indicates that our frequency-domain mixing strategy generally retains the label distribution of the original sample, thereby preserving the semantics of the augmented sample to a large extent.
}

\section{Experiments} \label{sec:experiments}
In this section, we conduct extensive experiments on multiple datasets. We evaluate the performance of our method on OOD generalization using both artificial and real-world data. We perform ablation studies to determine the contribution of each component. Moreover, we assess our method's impact on ID generalization. Parameter sensitivity experiments of the mixing ratio and window size can be found in Appendix \ref{app:parameter sensitivity}. 

\subsection{Experimental Setup}  \label{sec:experimental setup}
\subsubsection{Datasets and Evaluation Metrics}
\begin{table*}[htbp]
\centering
\small
\setlength{\aboverulesep}{0.3ex}
\setlength{\belowrulesep}{0.2ex}
\setlength{\tabcolsep}{6pt}
\caption{Partition and reorganization of datasets.}
\vspace{-2mm}
\label{tab:partition_reorganization}
\renewcommand{\arraystretch}{0.9}
\begin{tabular}{cclll}
\toprule
\makecell{\textbf{Original} \textbf{Dataset}} & \textbf{Dataset} & \textbf{Partition Criteria} & \textbf{Training Set} & \textbf{Test Set} \\
\midrule
\multirow[c]{3}{*}[-12pt]{CIC-DoH} & $\mathcal{D}_1$ & Browser & \makecell[l]{\textbf{B}: Firefox \\ \textbf{M}: Iodine} & \makecell[l]{\textbf{B}: Chrome \\ \textbf{M}: dns2tcp, dnscat2} \\
\cmidrule{2-5}
& $\mathcal{D}_2$ & DoH server & \makecell[l]{\textbf{B}: AdGuard, CloudFlare \\ \textbf{M}: Iodine, dnscat2} & \makecell[l]{\textbf{B}: Google DNS, Quad9 \\ \textbf{M}: dns2tcp} \\
\cmidrule{2-5}
& $\mathcal{D}_3$ & \makecell[l]{Mixed partition: \\ the training set \\ contains traffic from \\ multiple browsers \\ and DoH servers} & \makecell[l]{\textbf{B}: Chrome\_AdGuard, \\Chrome\_CloudFlare, \\ Firefox\_Google DNS,\\ Firefox\_Quad9 \\ \textbf{M}: dns2tcp} & \makecell[l]{\textbf{B}:  Chrome\_Google DNS, \\Chrome\_Quad9, \\ Firefox\_AdGuard, \\Firefox\_CloudFlare \\ \textbf{M}: Iodine, dnscat2} \\
\midrule
\makecell[c]{CIC-IDS2017, \\ CIC-IDS2018} & $\mathcal{D}_4$ & \makecell[l]{Different network \\ environments} & \makecell[l]{July 6th data in CIC-IDS2017 \\ (\textbf{M}: web attack, infiltration)} & \makecell[l]{February 22nd data in CIC-IDS2018 \\(\textbf{M}: web attack)} \\
\midrule
MAWILab & $\mathcal{D}_5$ & Date & Data collected in May 12th, 2024  & Data collected in May 19th, 2024 \\
\midrule
CSD & $\mathcal{D}_6$ & Date & Data collected in June 2nd, 2024 & Data collected in August 19th, 2024 \\
\bottomrule
\end{tabular}
\begin{tablenotes}
\setlength{\leftskip}{2.3em}
\item * \textbf{B}: Benign traffic. \textbf{M}: Malicious traffic. 
\end{tablenotes}
\renewcommand{\arraystretch}{1.0}
\setlength{\aboverulesep}{0.6ex}
\setlength{\belowrulesep}{0.4ex}
\setlength{\tabcolsep}{6pt}
\vspace{-1.5mm}
\end{table*}

We use multiple datasets to evaluate our method, including CIRA-CIC-DoHBrw2020 \cite{9251211} (CIC-DoH), CIC-IDS2017 \cite{SharafaldinLG18} and CSE-CIC-IDS2018 \cite{cicids2018} (CIC-IDS2018). Since these datasets were collected in artificial environments, we additionally employ two real-world datasets for evaluation in practical scenarios: the MAWILab dataset \cite{mawilab} and a real-world dataset (called CSD) provided by our collaborators. These datasets are introduced as follows:
\begin{enumerate}[left=0pt,label=$\bullet$]
    \item CIRA-CIC-DoHBrw2020: A malicious traffic detection dataset collected by the Canadian Institute for Cybersecurity (CIC). It contains a large amount of benign traffic and encrypted malicious traffic based on DoH (DNS over HTTPS). 
    \item CIC-IDS2017: An intrusion detection dataset collected by the CIC in a controlled environment. It includes various types of malicious attacks, such as botnet, DDoS, and infiltration.
    \item CSE-CIC-IDS2018: An intrusion detection dataset collected by the CIC. It is similar to CIC-IDS2017 in terms of basic characteristics; however, it has a larger network scale and more diverse traffic.
    \item MAWILab: This dataset is collected by the WIDE project \cite{wide}. It is a real-world dataset that contains a large amount of backbone traffic and labels of traffic anomalies obtained using an advanced graph-based methodology.
    \item CSD: This is our private dataset collected in real-world network environments. It includes extensive benign traffic, as well as malicious traffic from several monitored suspicious IP addresses. 
\end{enumerate}

To meet the needs of evaluating OOD generalization, we partition and reorganize these datasets.
In the CIRA-CIC-DoHBrw2020 dataset, benign traffic is generated by pairing two browsers (Chrome and Firefox) with four DoH servers (AdGuard, Cloudflare, Google DNS, and Quad9), yielding eight combinations in total. Malicious traffic is generated by three tools: dns2tcp, dnscat2, and Iodine. In real-world scenarios, users may favor a specific browser or DoH server. Consequently, data generated by other browsers or servers would become ``out-of-distribution data''.
Accordingly, we partition the combinations within the CIRA-CIC-DoHBrw2020 dataset to create three reorganized datasets for experiments, denoted as $\mathcal{D}_1\sim\mathcal{D}_3$. For benign traffic, $\mathcal{D}_1$ and $\mathcal{D}_2$ are partitioned according to browser types and DoH servers, respectively, while $\mathcal{D}_3$ adopts a mixed partitioning. Since malicious traffic is generated by three tunneling tools, we select traffic from one or two of these tools to include in the training set each time, while the remaining traffic is assigned to the test set.

In addition, we let the models train on a subset of CIC-IDS2017 and evaluate on a subset of CIC-IDS2018 (this reorganized dataset is denoted as $\mathcal{D}_4$). The training data and test data of $\mathcal{D}_4$ come from almost entirely different environments. Their only similarity is that the types of malicious traffic are approximately the same, thus making it more challenging.  For the real-world datasets (denoted as $\mathcal{D}_5\sim\mathcal{D}_6$), we use data collected on one day for training, while data collected on another day serve as test data. Under our data partitioning strategy, the OOD scenarios described in Section \ref{sec:introduction} are naturally reflected. For example, dataset $\mathcal{D}_1$ includes variants where the same DoH server is paired with different browsers; for real-world datasets, although data are split by date, they are collected from the same monitoring nodes and therefore may still contain similar flows or flow variants. A detailed summary of all datasets is provided in Table \ref{tab:partition_reorganization}. In Table \ref{tab:partition_reorganization}, ``Chrome\_AdGuard'' denotes traffic generated by the combination of Chrome and AdGuard, while ``Chrome'' refers to all traffic involving Chrome in the combinations (i.e., including all four combinations involving Chrome).

We conduct binary classification on the flows within these datasets to detect malicious traffic. We divide the training sets in Table \ref{tab:partition_reorganization} into training, validation, and in-distribution test sets at a ratio of 6:2:2, which allows us to assess the impact of our data augmentation method on the ID data. The test sets in Table \ref{tab:partition_reorganization} are used to evaluate the OOD generalization of the model. As these datasets are imbalanced, we adopt precision, recall, F1 score (F1), and false positive rate (FPR) as our evaluation metrics. Due to space constraints, we report the F1 score (as it provides a balanced assessment of both precision and recall) and the FPR in Section \ref{sec:experiments}. Detailed results including all metrics can be found in Appendix \ref{app:detailed results}.

\subsubsection{Models and Baselines}

\noindent \textbf{Models.} We select several models for traffic classification, including three widely used general models and seven specialized models, to evaluate the augmentation effects of our method. The general models are LSTM \cite{LSTM6795963}, 1D-CNN \cite{726791}, and Transformer \cite{10.5555/3295222.3295349}. These three models are commonly used as fundamental architectures in network traffic classification tasks \cite{LiuHXCL19,SirinamMR019,HanCLZJL23}. The specialized models are:
\begin{enumerate}[left=0pt,label=$\bullet$]
    \item DF \cite{10.1145/3243734.3243768}: This model employs a carefully designed architecture based on stacked 1D-CNNs for traffic classification.
    \item BAPM \cite{GuanXGLCL21}: This model extracts features through convolutional layers, divides the representations into blocks, and employs a multihead attention mechanism to fuse the information of these blocks for traffic classification.
    \item TMWF \cite{JinLLS23}: This model uses the backbone of DF to extract sequential features and then applies a Transformer-based architecture for traffic classification.
    \item LUCID \cite{LUCID}: This model extracts multiple packet attributes as sequential features and structures them into matrices, employing a 2D-CNN for malicious traffic detection.
    \item Whisper \cite{10.1145/3460120.3484585}: This model extracts frequency-domain features for malicious traffic detection and is considered one of the state-of-the-art approaches. While it is originally unsupervised, we employ the supervised implementation by Yan et al. \cite{CertTA} to leverage label information.
    \item MATEC \cite{MATEC}: This model integrates the multi-head attention mechanism with CNNs for traffic classification.
    \item SmartDetector \cite{SAM}: A recently proposed model that transforms packet sequence features into word embedding sequences and incorporates contrastive learning for malicious traffic detection.
\end{enumerate}

\noindent \textbf{Features. }Considering that the packet length sequence is a widely adopted and effective feature \cite{LiuHXCL19,XieCDX00SZ23,QingYDCL000024,JiangCLGXL23,FauvelC023}, we employ it for all three generic models. For specialized models, we primarily follow the feature sets used in their original papers (but may make modifications to suit our scenario). In detail, DF, BAPM, and TMWF use packet length sequences; MATEC and SmartDetector use packet length, IAT, and packet direction sequences; Whisper uses packet length and IAT sequences; and LUCID follows the same feature sets as in \cite{LUCID}. We apply the augmentations to packet length and IAT (if available) sequences; for LUCID, the augmentations are applied to packet length, TCP payload length, and UDP payload length sequences. Additionally, for LUCID on datasets $\mathcal{D}_1\sim\mathcal{D}_3$, we exclude three features---relative time, frame protocols, and TCP window size---to prevent shortcut exploitation (see Section \ref{sec:case study}).

\begin{table*}[htbp]
\centering
\caption{Comparison of different data augmentation methods on artificial datasets.}
\vspace{-2mm}
\label{tab:method_compare}
\setlength{\tabcolsep}{2.5pt} 
\renewcommand{\arraystretch}{0.65}
\resizebox{0.975\textwidth}{!}{
\begin{tabular}{cc cccccccc cccccccc}
\toprule
\multirow{2}{*}{\textbf{Model}} & \multirow{2}{*}{\textbf{Dataset}} & \multicolumn{2}{c}{\textbf{NoAug}} & \multicolumn{2}{c}{\textbf{Jittering}} & \multicolumn{2}{c}{\textbf{Permutation}} & \multicolumn{2}{c}{\textbf{RM}} & \multicolumn{2}{c}{\textbf{Mixup}} & \multicolumn{2}{c}{\textbf{FreqMix}} & \multicolumn{2}{c}{\textbf{Rosetta}} & \multicolumn{2}{c}{\textbf{Ours}} \\
\cmidrule(lr){3-4} \cmidrule(lr){5-6} \cmidrule(lr){7-8} \cmidrule(lr){9-10} \cmidrule(lr){11-12} \cmidrule(lr){13-14} \cmidrule(lr){15-16} \cmidrule(lr){17-18}
& & F1\,$\uparrow$ & FPR\,$\downarrow$ & F1\,$\uparrow$ & FPR\,$\downarrow$ & F1\,$\uparrow$ & FPR\,$\downarrow$ & F1\,$\uparrow$ & FPR\,$\downarrow$ & F1\,$\uparrow$ & FPR\,$\downarrow$ & F1\,$\uparrow$ & FPR\,$\downarrow$ & F1\,$\uparrow$ & FPR\,$\downarrow$ & F1\,$\uparrow$ & FPR\,$\downarrow$ \\
\midrule

\multirow{4}{*}{\textbf{LSTM}} 
& $\mathcal{D}_1$ & 0.7429 & \textbf{0.0018} & 0.7331 & \underline{0.0019} & 0.6954 & 0.0025 & 0.6951 & 0.0028 & 0.8752 & 0.0136 & \underline{0.9259} & 0.0134 & 0.6582 & 0.0129 & \textbf{0.9528} & 0.0171 \\
& $\mathcal{D}_2$ & 0.3865 & \underline{0.0006} & 0.6724 & 0.0011 & 0.5567 & \textbf{0.0002} & 0.4692 & 0.0011 & \underline{0.9039} & 0.0027 & 0.8588 & 0.0032 & 0.5623 & 0.0125 & \textbf{0.9182} & 0.0045 \\
& $\mathcal{D}_3$ & 0.5794 & \textbf{0.0001} & 0.5935 & \textbf{0.0001} & \underline{0.7996} & \underline{0.0002} & 0.6126 & 0.0012 & 0.6943 & 0.0006 & 0.7252 & 0.0012 & 0.7534 & 0.0148 & \textbf{0.8149} & 0.0035 \\
& $\mathcal{D}_4$ & 0 & -- & 0 & -- & 0 & -- & 0 & -- & 0 & -- & 0 & -- & 0 & -- & 0 & -- \\
\midrule

\multirow{4}{*}{\textbf{1D-CNN}} 
& $\mathcal{D}_1$ & 0.7328 & \textbf{0.0056} & 0.7505 & \underline{0.0062} & 0.7563 & 0.0080 & 0.7328 & 0.0097 & \underline{0.7779} & 0.0183 & 0.7158 & 0.0187 & 0.7503 & 0.0165 & \textbf{0.8001} & 0.0358 \\
& $\mathcal{D}_2$ & 0.6047 & \textbf{0.0063} & 0.6073 & \underline{0.0064} & 0.6003 & 0.0082 & 0.5945 & 0.0130 & 0.6042 & 0.0153 & \underline{0.6118} & 0.0145 & 0.5680 & 0.0257 & \textbf{0.6533} & 0.0571 \\
& $\mathcal{D}_3$ & 0.8202 & \textbf{0.0041} & 0.7702 & \underline{0.0046} & 0.7477 & 0.0057 & 0.8276 & 0.0091 & 0.8682 & 0.0149 & \underline{0.9263} & 0.0098 & 0.8318 & 0.0288 & \textbf{0.9301} & 0.0253 \\
& $\mathcal{D}_4$ & 0.0109 & 0.0013 & 0.0114 & \textbf{0.0007} & 0 & -- & 0.0110 & 0.0011 & 0.0112 & 0.0009 & 0.0112 & \underline{0.0008} & \underline{0.5126} & \underline{0.0008} & \textbf{0.5714} & 0.0011 \\
\midrule

\multirow{4}{*}{\textbf{Transformer}} 
& $\mathcal{D}_1$ & 0.7405 & 0.0060 & 0.7399 & 0.0065 & 0.6677 & \textbf{0.0039} & 0.7210 & 0.0246 & 0.8026 & \underline{0.0050} & \underline{0.8578} & 0.0165 & 0.6288 & 0.0106 & \textbf{0.9014} & 0.0201 \\
& $\mathcal{D}_2$ & 0.5692 & \textbf{0.0039} & 0.5690 & 0.0043 & 0.6114 & \underline{0.0040} & 0.6266 & 0.0063 & 0.7559 & 0.0047 & \underline{0.8005} & 0.0074 & 0.5827 & 0.0157 & \textbf{0.9144} & 0.0111 \\
& $\mathcal{D}_3$ & 0.7059 & 0.0012 & 0.6734 & 0.0013 & 0.5659 & \underline{0.0007} & 0.6027 & 0.0018 & 0.7530 & \textbf{0.0005} & 0.7602 & 0.0022 & \underline{0.7808} & 0.0145 & \textbf{0.7818} & 0.0021 \\
& $\mathcal{D}_4$ & 0 & -- & 0 & -- & 0 & -- & 0 & -- & 0 & -- & 0 & -- & \underline{0.3306} & \textbf{0.0105} & \textbf{0.3799} & \underline{0.0108} \\
\midrule

\multirow{4}{*}{\textbf{DF}} 
& $\mathcal{D}_1$ & 0.7156 & 0.0109 & 0.7228 & 0.0086 & \underline{0.7703} & \underline{0.0038} & 0.6851 & \underline{0.0038} & 0.6975 & \textbf{0.0023} & 0.7513 & 0.0074 & 0.6256 & 0.0122 & \textbf{0.8210} & 0.0071 \\
& $\mathcal{D}_2$ & 0.7342 & 0.0028 & 0.5772 & \underline{0.0025} & \underline{0.7800} & \textbf{0.0024} & 0.6349 & 0.0067 & 0.7344 & \textbf{0.0024} & 0.6085 & 0.0035 & 0.5895 & 0.0101 & \textbf{0.8446} & 0.0080 \\
& $\mathcal{D}_3$ & 0.6379 & \textbf{0.0003} & 0.6390 & \underline{0.0005} & 0.5770 & 0.0007 & 0.6806 & 0.0011 & 0.6856 & \textbf{0.0003} & 0.7002 & 0.0024 & \textbf{0.7550} & 0.0115 & \underline{0.7226} & 0.0007 \\
& $\mathcal{D}_4$ & 0 & -- & 0 & -- & 0 & -- & 0 & -- & 0 & -- & 0 & -- & \textbf{0.2837} & \textbf{0.0001} & \underline{0.0854} & \underline{0.0919} \\
\midrule

\multirow{4}{*}{\textbf{BAPM}} 
& $\mathcal{D}_1$ & 0.7058 & 0.0084 & 0.7310 & 0.0084 & 0.7336 & 0.0109 & 0.7623 & \textbf{0.0047} & \underline{0.7755} & \underline{0.0051} & 0.7416 & 0.0662 & 0.6814 & 0.0125 & \textbf{0.8163} & 0.0495 \\
& $\mathcal{D}_2$ & 0.5783 & 0.0051 & 0.5064 & \textbf{0.0032} & 0.5675 & 0.0112 & 0.5734 & 0.0178 & 0.6902 & \underline{0.0041} & \underline{0.7022} & 0.0124 & 0.5909 & 0.0153 & \textbf{0.7054} & 0.0523 \\
& $\mathcal{D}_3$ & 0.6764 & \textbf{0.0008} & 0.6414 & \textbf{0.0008} & 0.5689 & 0.0014 & 0.6422 & 0.0014 & 0.7039 & \underline{0.0009} & 0.6444 & 0.0034 & \underline{0.7982} & 0.0190 & \textbf{0.8121} & 0.0238 \\
& $\mathcal{D}_4$ & 0.1390 & 0.0498 & 0.0062 & \textbf{0.0120} & \underline{0.2616} & 0.0188 & 0.1280 & 0.0555 & 0.0052 & \underline{0.0167} & 0 & -- & 0.2230 & 0.0245 & \textbf{0.3833} & 0.0302 \\
\midrule

\multirow{4}{*}{\textbf{TMWF}} 
& $\mathcal{D}_1$ & \underline{0.8323} & 0.0162 & 0.7865 & 0.0042 & 0.7280 & \underline{0.0026} & 0.7131 & 0.0028 & 0.7735 & \textbf{0.0015} & 0.8003 & 0.0071 & 0.6084 & 0.0150 & \textbf{0.9059} & 0.0166 \\
& $\mathcal{D}_2$ & 0.7617 & \underline{0.0018} & 0.6949 & 0.0029 & \underline{0.8199} & 0.0023 & 0.6874 & 0.0047 & 0.7308 & \textbf{0.0009} & 0.7819 & 0.0032 & 0.5550 & 0.0090 & \textbf{0.8366} & 0.0036 \\
& $\mathcal{D}_3$ & 0.6815 & 0.0003 & 0.6654 & \underline{0.0002} & 0.6493 & 0.0006 & 0.6862 & 0.0015 & 0.7138 & \textbf{0.0001} & 0.6948 & 0.0007 & \textbf{0.7658} & 0.0115 & \underline{0.7463} & 0.0014 \\
& $\mathcal{D}_4$ & 0 & -- & 0 & -- & 0 & -- & 0 & -- & 0 & -- & 0 & - & 0 & -- & \textbf{0.4963} & \textbf{0.0183} \\
\midrule

\multirow{4}{*}{\textbf{LUCID}} 
& $\mathcal{D}_1$ & 0.3247 & 0.0116 & 0.3317 & 0.0090 & 0.3090 & 0.0092 & 0.3306 & 0.0100 & 0.3185 & \textbf{0.0072} & \underline{0.4157} & 0.0176 & / & / & \textbf{0.4285} & \underline{0.0082} \\
& $\mathcal{D}_2$ & \underline{0.3104} & 0.0150 & 0.3077 & 0.0097 & 0.1653 & \textbf{0.0060} & 0.3063 & 0.0165 & 0.2704 & 0.0092 & 0.2064 & \underline{0.0062} & / & / & \textbf{0.3973} & 0.0093 \\
& $\mathcal{D}_3$ & 0.7896 & 0.0115 & 0.7493 & 0.0185 & 0.7464 & 0.0385 & 0.7924 & 0.0178 & 0.8212 & \textbf{0.0059} & \underline{0.8215} & \underline{0.0062} & / & / & \textbf{0.8245} & 0.0071 \\
& $\mathcal{D}_4$ & 0 & -- & 0 & -- & 0 & -- & 0 & -- & 0 & -- & 0 & -- & / & / & 0 & -- \\
\midrule

\multirow{4}{*}{\textbf{Whisper}} 
& $\mathcal{D}_1$ & 0.7211 & 0.0043 & 0.2843 & \underline{0.0027} & 0.2999 & 0.0040 & \underline{0.7909} & 0.0044 & 0.3001 & \textbf{0.0011} & 0.4157 & 0.0176 & / & / & \textbf{0.8222} & 0.0109 \\
& $\mathcal{D}_2$ & 0.7553 & 0.0031 & 0.7465 & \textbf{0.0022} & 0.7320 & 0.0031 & \underline{0.7603} & 0.0045 & 0.0876 & \underline{0.0030} & 0.2578 & 0.0075 & / & / & \textbf{0.7969} & 0.0086 \\
& $\mathcal{D}_3$ & 0.8173 & \textbf{0.0042} & \underline{0.8495} & 0.0054 & 0.8470 & 0.0063 & 0.7854 & 0.0194 & 0.7894 & \underline{0.0045} & 0.8231 & 0.0201 & / & / & \textbf{0.9077} & 0.0240 \\
& $\mathcal{D}_4$ & 0 & -- & 0 & -- & 0 & -- & 0 & -- & 0 & -- & 0 & -- & / & / & 0 & -- \\
\midrule

\multirow{4}{*}{\textbf{MATEC}} 
& $\mathcal{D}_1$ & 0.2974 & 0.1108 & 0.2937 & 0.1115 & 0.2728 & \underline{0.1106} & 0.2668 & 0.1651 & 0.3148 & 0.1624 & \underline{0.4158} & \textbf{0.0887} & / & / & \textbf{0.8312} & 0.1635 \\
& $\mathcal{D}_2$ & 0.1469 & 0.0015 & 0.1458 & 0.0017 & 0.1584 & 0.0028 & 0.1350 & 0.0026 & 0.2157 & \textbf{0.0010} & \underline{0.4080} & 0.0027 & / & / & \textbf{0.9748} & \underline{0.0014} \\
& $\mathcal{D}_3$ & 0.8909 & 0.0008 & 0.9175 & 0.0041 & 0.7287 & 0.0051 & 0.8430 & 0.0078 & \underline{0.9248} & \textbf{0.0004} & 0.8967 & 0.0030 & / & / & \textbf{0.9379} & \underline{0.0005} \\
& $\mathcal{D}_4$ & 0.3396 & \textbf{0.0008} & 0.3333 & 0.0011 & \underline{0.3915} & 0.0018 & 0.3349 & \underline{0.0010} & 0.2939 & 0.0033 & 0.3318 & 0.0011 & / & / & \textbf{0.5199} & 0.0030 \\
\midrule

\multirow{4}{*}{\textbf{SmartDetector}} 
& $\mathcal{D}_1$ & 0.9935 & 0.0017 & 0.7209 & 0.0781 & 0.9928 & \textbf{0.0002} & \underline{0.9951} & 0.0029 & 0.9557 & 0.0024 & 0.9677 & 0.0074 & / & / & \textbf{0.9992} & \underline{0.0006} \\
& $\mathcal{D}_2$ & \underline{0.9977} & \textbf{0} & 0.8396 & \underline{0.0003} & 0.9931 & \textbf{0} & 0.9877 & \textbf{0} & 0.9627 & \underline{0.0003} & 0.9698 & \textbf{0} & / & / & \textbf{0.9999} & \textbf{0} \\
& $\mathcal{D}_3$ & 0.9994 & 0.0002 & 0.9823 & \underline{0.0001} & \textbf{0.9999} & \textbf{0} & \underline{0.9998} & \textbf{0} & 0.9673 & \textbf{0} & 0.9957 & 0.0002 & / & / & 0.9996 & \underline{0.0001} \\
& $\mathcal{D}_4$ & 0 & -- & \textbf{0.5353} & \textbf{0.0024} & 0 & -- & 0 & -- & 0.3820 & 0.0106 & 0.2618 & 0.0238 & / & / & \underline{0.4260} & \underline{0.0076} \\

\bottomrule

\end{tabular}
}
\begin{tablenotes}
\small
\setlength{\leftskip}{-1em}
\setlength{\rightskip}{1em}
\item * The best results among the different methods are marked in \textbf{bold}, and the second-best results are \underline{underlined}. We mark the FPR as ``--'' when the F1 score is 0 (as FPR is meaningless in this case). Rosetta cannot be applied to models utilizing features other than packet length, so we mark this case as ``/''. 
\end{tablenotes}
\vspace{-1.5mm}
\end{table*}

\noindent \textbf{Baselines. }We select six data augmentation methods for comparison. These include two conventional time series augmentation techniques: \textit{Jittering} \cite{IglesiasTGMC23}, which adds noise to the data, and \textit{Permutation} \cite{IglesiasTGMC23}, which shuffles sequence segments. We also compare against two general-purpose methods: \textit{Random Masking (RM)} \cite{DevlinCLT19} and \textit{Mixup} \cite{ZhangCDL18}. Random masking randomly masks elements in the sequence, and Mixup performs linear mixing on two samples and their labels. In addition, we include \textit{Standard Frequency-Domain Mixing (FreqMix)} \cite{KimHK21,abs-2302-09292} described in Section \ref{sec:Formal Definition}, which mixes the frequency-domain features of time series data. Finally, we adopt an augmentation method designed for traffic classification, \textit{Rosetta} \cite{XieCDX00SZ23}. Rosetta employs contrastive learning to minimize the distance between a packet length sequence and its augmented version, thereby training a robust feature extractor to obtain more stable representations. We denote our augmentation method as ``\textit{Ours}'' and the case without data augmentation as``\textit{NoAug}''.

More details, including the implementation of pre-augmentation and parameter settings, can be found in Appendix  \ref{app:experimental details}.


\subsection{Evaluation in Artificial Environment} \label{sec:augment performance}
We first evaluate the effectiveness of our method in improving the model’s OOD generalization in artificial environments ($\mathcal{D}_1\sim\mathcal{D}_4$), and compare it with other augmentation methods. Data collected in artificial environments have more reliable labels and less noise. The results are presented in Table \ref{tab:method_compare}. More detailed metrics for our method can be found in Table \ref{tab:ood_results_full} in Appendix \ref{app:detailed results}.

\begin{table*}[htbp]
\centering
\caption{Comparison of different data augmentation methods on real-world datasets.}
\vspace{-2mm}
\label{tab:real-world compare}
\setlength{\tabcolsep}{2.0pt} 
\renewcommand{\arraystretch}{0.65}
\resizebox{0.96\textwidth}{!}{
\begin{tabular}{cc cccccccc cccccccc}
\toprule
\multirow{2}{*}{\textbf{Model}} & \multirow{2}{*}{\textbf{Dataset}} & \multicolumn{2}{c}{\textbf{NoAug}} & \multicolumn{2}{c}{\textbf{Jittering}} & \multicolumn{2}{c}{\textbf{Permutation}} & \multicolumn{2}{c}{\textbf{RM}} & \multicolumn{2}{c}{\textbf{Mixup}} & \multicolumn{2}{c}{\textbf{FreqMix}} & \multicolumn{2}{c}{\textbf{Rosetta}} & \multicolumn{2}{c}{\textbf{Ours}} \\
\cmidrule(lr){3-4} \cmidrule(lr){5-6} \cmidrule(lr){7-8} \cmidrule(lr){9-10} \cmidrule(lr){11-12} \cmidrule(lr){13-14} \cmidrule(lr){15-16} \cmidrule(lr){17-18}
& & F1\,$\uparrow$ & FPR\,$\downarrow$ & F1\,$\uparrow$ & FPR\,$\downarrow$ & F1\,$\uparrow$ & FPR\,$\downarrow$ & F1\,$\uparrow$ & FPR\,$\downarrow$ & F1\,$\uparrow$ & FPR\,$\downarrow$ & F1\,$\uparrow$ & FPR\,$\downarrow$ & F1\,$\uparrow$ & FPR\,$\downarrow$ & F1\,$\uparrow$ & FPR\,$\downarrow$ \\
\midrule

\multirow{2}{*}{\textbf{LSTM}} 
& $\mathcal{D}_5$ & 0.7657 & 0.1831 & 0.7295 & 0.1987 & 0.7249 & 0.2096 & 0.7495 & 0.2117 & \underline{0.7674} & \underline{0.1737} & 0.7534 & 0.1785 & 0.6139 & 0.2439 & \textbf{0.7677} & \textbf{0.1597} \\
& $\mathcal{D}_6$ & 0.7499 & 0.1574 & 0.6877 & \textbf{0.1116} & 0.6994 & 0.1665 & 0.7579 & 0.1485 & 0.7521 & 0.1337 & \underline{0.7596} & 0.1429 & 0.5664 & 0.2764 & \textbf{0.7623} & \underline{0.1320} \\
\midrule

\multirow{2}{*}{\textbf{1D-CNN}} 
& $\mathcal{D}_5$ & 0.5701 & \textbf{0.1763} & \textbf{0.5732} & 0.1985 & 0.5142 & 0.2778 & 0.5590 & \underline{0.1968} & \underline{0.5706} & 0.2334 & 0.5518 & 0.2080 & 0.5569 & 0.2551 & 0.5641 & 0.2523 \\
& $\mathcal{D}_6$ & \textbf{0.6240} & \underline{0.2457} & 0.5967 & 0.2788 & 0.4969 & 0.3522 & 0.5589 & \textbf{0.2184} & 0.6043 & 0.2438 & 0.5982 & 0.2507 & 0.4309 & 0.2745 & \underline{0.6199} & 0.3293 \\
\midrule

\multirow{2}{*}{\textbf{Transformer}} 
& $\mathcal{D}_5$ & 0.6983 & \textbf{0.1929} & 0.6884 & \underline{0.1989} & 0.6757 & 0.2263 & 0.7089 & 0.2007 & \underline{0.7168} & 0.2042 & 0.6936 & 0.2412 & 0.6354 & 0.2562 & \textbf{0.7270} & 0.2122 \\
& $\mathcal{D}_6$ & 0.6780 & \textbf{0.1599} & 0.6459 & 0.2104 & 0.6045 & 0.2418 & \underline{0.6933} & \underline{0.1643} & 0.6922 & 0.1697 & 0.6724 & 0.1704 & 0.5794 & 0.2676 & \textbf{0.7027} & 0.1992 \\
\midrule

\multirow{2}{*}{\textbf{DF}} 
& $\mathcal{D}_5$ & 0.7063 & \underline{0.1825} & 0.7157 & 0.1900 & 0.6916 & 0.2415 & 0.7015 & 0.2085 & \underline{0.7296} & 0.2072 & 0.7205 & 0.2057 & 0.5785 & 0.2479 & \textbf{0.7387} & \textbf{0.1733} \\
& $\mathcal{D}_6$ & \underline{0.7499} & 0.1415 & 0.7140 & 0.2081 & 0.6569 & 0.1759 & 0.6607 & \textbf{0.1191} & 0.7439 & 0.1476 & 0.7190 & 0.1476 & 0.5392 & 0.2463 & \textbf{0.7523} & \underline{0.1343} \\
\midrule

\multirow{2}{*}{\textbf{BAPM}} 
& $\mathcal{D}_5$ & \underline{0.6747} & 0.2394 & 0.6637 & \underline{0.2151} & 0.6199 & 0.3113 & 0.6610 & 0.2340 & 0.6667 & 0.2493 & 0.6669 & 0.2613 & 0.6093 & 0.2683 & \textbf{0.6881} & \textbf{0.2073} \\
& $\mathcal{D}_6$ & \underline{0.7204} & 0.1662 & 0.6638 & 0.2657 & 0.5725 & 0.3190 & 0.6621 & \textbf{0.1550} & 0.7143 & 0.1742 & 0.6918 & \underline{0.1567} & 0.4996 & 0.2057 & \textbf{0.7285} & 0.1846 \\
\midrule

\multirow{2}{*}{\textbf{TMWF}} 
& $\mathcal{D}_5$ & \underline{0.7292} & 0.1866 & 0.7079 & \underline{0.1826} & 0.6951 & 0.2336 & 0.7170 & 0.1903 & 0.7266 & 0.1888 & 0.7258 & 0.2026 & 0.5415 & \textbf{0.1287} & \textbf{0.7320} & 0.1848 \\
& $\mathcal{D}_6$ & 0.7276 & 0.1387 & 0.6861 & 0.1584 & 0.6331 & 0.1589 & 0.7202 & 0.1319 & \underline{0.7306} & \underline{0.1307} & 0.7312 & \textbf{0.1292} & 0.5106 & 0.2517 & \textbf{0.7393} & 0.1412 \\
\midrule

\multirow{2}{*}{\textbf{LUCID}} 
& $\mathcal{D}_5$ & 0 & -- & 0 & -- & 0 & -- & 0 & -- & 0 & -- & 0 & -- & / & / & 0 & -- \\
& $\mathcal{D}_6$ & \textbf{0.3593} & 0.0496 & 0.2932 & \underline{0.0358} & 0.2884 & 0.0365 & 0.3099 & 0.0385 & 0.2778 & 0.0359 & 0.1844 & \textbf{0.0179} & / & / & \underline{0.3378} & 0.0446 \\
\midrule

\multirow{2}{*}{\textbf{Whisper}} 
& $\mathcal{D}_5$ & 0.4315 & 0.2024 & 0.3818 & 0.1959 & 0.3346 & 0.2657 & 0.4118 & 0.2464 & \underline{0.4644} & \textbf{0.1679} & 0.2518 & 1 & / & / & \textbf{0.4850} & \underline{0.1682} \\
& $\mathcal{D}_6$ & 0.2550 & 0.1052 & 0.1989 & \underline{0.0788} & 0.2328 & 0.0957 & 0.2274 & 0.0911 & \underline{0.2579} & 0.1090 & 0 & -- & / & / & \textbf{0.6338} & \textbf{0.0516} \\
\midrule

\multirow{2}{*}{\textbf{MATEC}} 
& $\mathcal{D}_5$ & 0.7024 & 0.2365 & 0.7145 & 0.2191 & 0.6358 & 0.2305 & 0.6954 & 0.2290 & \underline{0.7215} & \underline{0.2110} & 0.7177 & 0.2239 & / & / & \textbf{0.7247} & \textbf{0.2032} \\
& $\mathcal{D}_6$ & 0.6845 & \underline{0.2213} & 0.6607 & 0.1965 & 0.5396 & 0.4725 & 0.6607 & \textbf{0.1965} & \textbf{0.6936} & 0.2219 & \underline{0.6876} & 0.2403 & / & / & 0.6857 & 0.2241 \\
\midrule

\multirow{2}{*}{\textbf{SmartDetector}} 
& $\mathcal{D}_5$ & 0.3871 & 0.2675 & 0.2874 & 0.4854 & 0.3247 & 0.3331 & \underline{0.4074} & \underline{0.2451} & 0.3762 & 0.3264 & 0.3697 & 0.2925 & / & / & \textbf{0.4769} & \textbf{0.2052} \\
& $\mathcal{D}_6$ & 0.7735 & 0.0906 & 0.7770 & 0.1420 & 0.7546 & \textbf{0.0865} & 0.7604 & 0.1554 & \underline{0.7987} & \underline{0.0866} & 0.7425 & 0.1033 & / & / & \textbf{0.8157} & 0.0991 \\
\bottomrule
\end{tabular}
}
\begin{tablenotes}
\small
\setlength{\leftskip}{-1em}
\setlength{\rightskip}{1em}
\item * The best results among the different methods are marked in \textbf{bold}, and the second-best results are \underline{underlined}. We mark the FPR as ``--'' when the F1 score is 0 (as FPR is meaningless in this case). Rosetta cannot be applied to models utilizing features other than packet length, so we mark this case as ``/''. 
\end{tablenotes}
\vspace{-4mm}
\end{table*}

\begin{figure*}[htbp]
    \centering
    \includegraphics[width=0.85\linewidth]{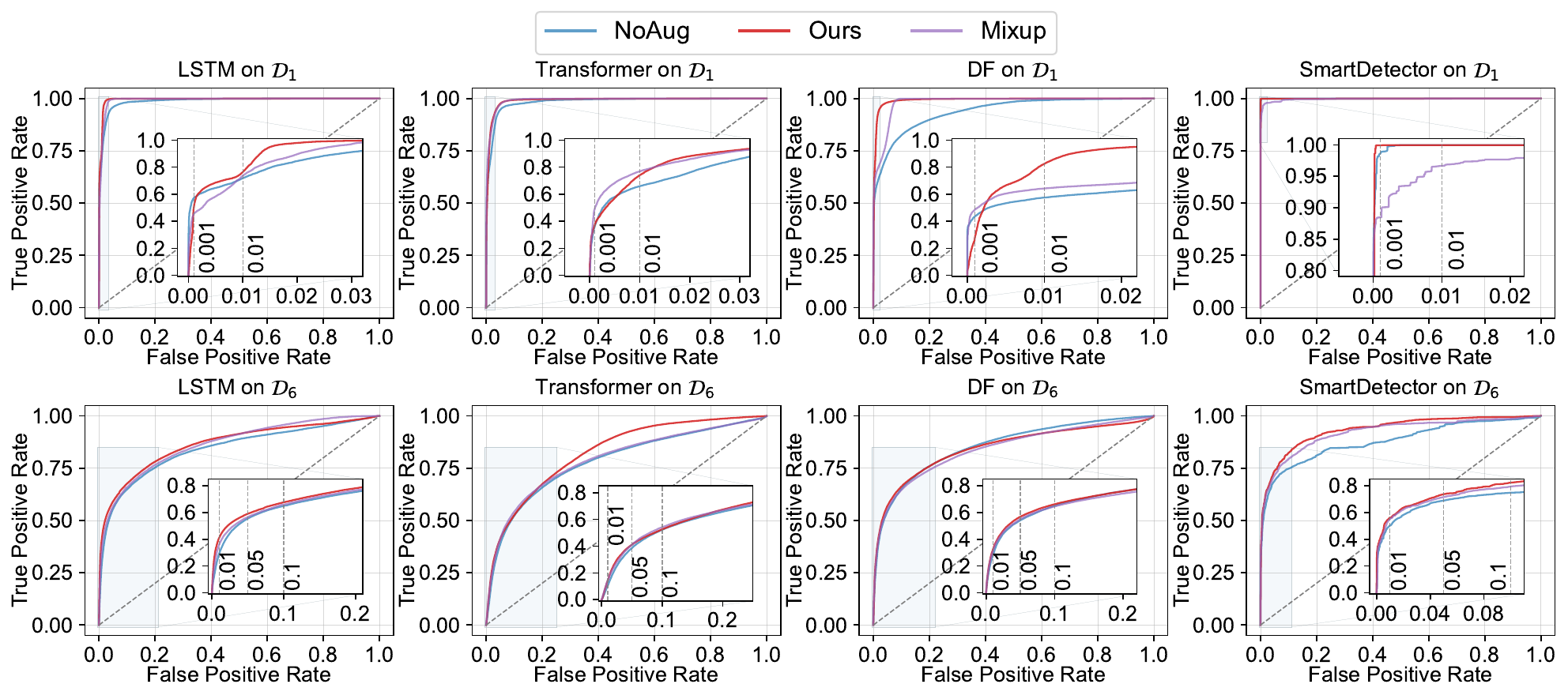}
    \captionsetup{skip=2pt}
    \caption{ROC curves of some representative models on the artificial dataset $\mathcal{D}_1$ (top row) and the real-world dataset $\mathcal{D}_6$ (bottom row). Insets show enlarged views of some key regions for detailed comparison.}
    \label{fig:roc_D1_D6}
    \vspace{-4mm}
\end{figure*}

As shown in Table \ref{tab:method_compare}, applying our data augmentation method typically improves the model's OOD generalization compared to training without augmentation. In many cases, the improvements are substantial, with the F1 score increasing by more than 10\%. On dataset $\mathcal{D}_2$ with the LSTM model, the improvement in F1 even exceeds 50\%. Table \ref{tab:method_compare} shows that our method leads to a slight increase in the FPR. Combined with the results in Table \ref{tab:ood_results_full}, it can be observed that our method usually enhances the recall by identifying more malicious samples, which consequently introduces some additional false positives. \rev{To further investigate the impact of our method on FPR, we select four representative models: the general sequence model LSTM, the general attention-based model Transformer, the classical CNN-based traffic classification model DF, and the state-of-the-art model SmartDetector. For each model, we plot the ROC curves of the models trained with our method (Ours) and with no augmentation (NoAug) on dataset $\mathcal{D}_1$ to analyze the true positive rate (TPR, i.e., recall) under different FPR levels. The results are shown in the top row of Figure \ref{fig:roc_D1_D6}. }

\begin{figure*}[htbp]
    \centering
    \includegraphics[width=\linewidth]{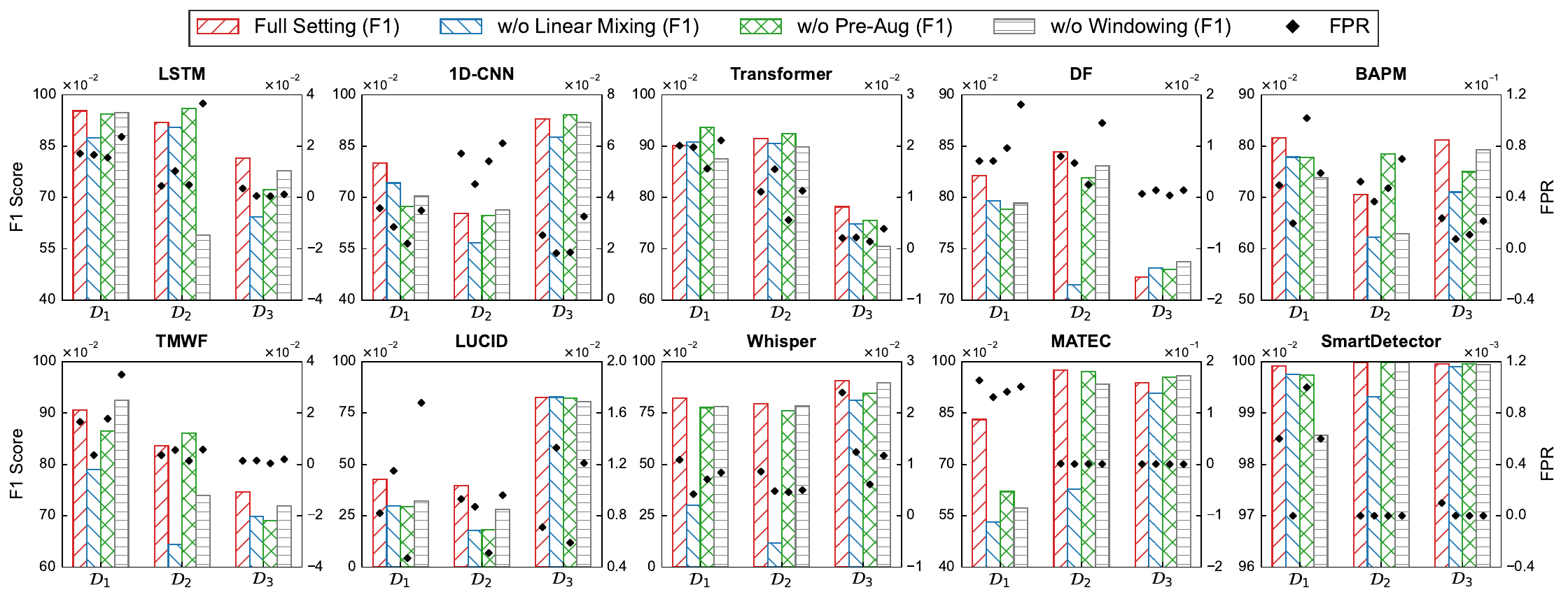}
    \vspace{-7mm}
    \caption{Results of the ablation study. For better visualization, the FPR axis may start from negative values.}
    \label{fig:ablation_study}
    \vspace{-4mm}
\end{figure*}

\rev{Compared with training without data augmentation, our method yields slightly lower TPR for LSTM, Transformer, and DF when the FPR is below approximately 0.1\%, 0.5\%, and 0.3\%, respectively. Once the allowable FPR exceeds these levels (as is often still acceptable in practical scenarios), our method gradually gains a clear advantage.  For SmartDetector, our method almost consistently outperforms training without augmentation at the same FPR level. }

Compared to other approaches, our method achieves the best F1 scores in most cases. For example, on dataset $\mathcal{D}_2$ with the Transformer model, our method achieves an F1 score of 0.9144, while the highest F1 score from any other method is 0.8005. Although our method shares some similarities with Mixup and FreqMix, it performs notably better than them. Moreover, compared to the non-augmented models, our method consistently improves the F1 score (except for some cases on $\mathcal{D}_4$ where all methods are ineffective), whereas other methods sometimes degrade performance. On dataset $\mathcal{D}_1$ with TMWF, Mixup and FreqMix reduce the F1 score by 5.88\% and 3.2\%, respectively, whereas our method boosts it by 7.36\%. Even for models like SmartDetector, which already achieves F1 scores of over 99\% without augmentation, our method can still enhance its F1 scores while simultaneously reducing the FPRs.
This suggests that our method is more suitable for network packet sequence data and that the generated new samples enable the models to better adapt to variations in network traffic. 
Rosetta outperforms our method in a few cases because it employs a feature extractor trained with contrastive learning \cite{XieCDX00SZ23}, whereas our method is non-learning-based. However, Rosetta is trained exclusively on packet length sequence and can only be applied to models that rely solely on packet length sequences as features, which significantly constrains its applicability.

\rev{To provide a more comprehensive evaluation of FPR, we also plot the ROC curves of Mixup in Figure \ref{fig:roc_D1_D6}, as Mixup achieves relatively strong performance in both F1 and FPR among all baselines. Overall, similar to the comparison with training without augmentation, Mixup may perform slightly better than our method at very low FPR levels (e.g., below approximately 0.1\% for LSTM), whereas our method shows an advantage in most other cases. Although Mixup performs better than our method on Transformer when the FPR is below approximately 1\%, this advantage disappears once the acceptable FPR exceeds this level. When a high TPR is required (e.g., 80\%), our method can usually achieve it with a lower FPR, significantly outperforming both Mixup and training without augmentation. }

$\mathcal{D}_4$ is a more challenging dataset, characterized by less training data and more severe distribution shifts. As a result, the original training set may not sufficiently support the model in learning robust features, often leading to the failure of non-augmented models. However, with our augmentation method, the detection capability can be partially recovered in many cases, whereas other methods typically remain ineffective or perform poorly. On TMWF, while all other methods fail, our approach improves the F1 score to 0.4963. For cases where all augmentations remain ineffective, we attribute this to the models' limited capabilities. Jittering performs well on SmartDetector in $\mathcal{D}_4$, which may be attributed to the possibility that features extracted from its augmented data happen to align with those of the OOD samples.

\begin{table*}[htbp]
\centering
\caption{Model performance on in-distribution data before and after applying our proposed data augmentation method.}
\vspace{-2mm}
\label{tab:augmentation_in_distribution}
\setlength{\tabcolsep}{2.5pt} 
\renewcommand{\arraystretch}{0.4}
\resizebox{0.824\textwidth}{!}{
\begin{tabular}{cc cccccccc cccccccc}
\toprule
\multirow{2}{*}{\textbf{Model}} & \multirow{2}{*}{\textbf{Aug}} & \multicolumn{2}{c}{$\mathcal{D}_1$ } & \multicolumn{2}{c}{$\mathcal{D}_2$} & \multicolumn{2}{c}{$\mathcal{D}_3$} & \multicolumn{2}{c}{$\mathcal{D}_4$} & \multicolumn{2}{c}{$\mathcal{D}_5$} & \multicolumn{2}{c}{$\mathcal{D}_6$}  \\
\cmidrule(lr){3-4} \cmidrule(lr){5-6} \cmidrule(lr){7-8} \cmidrule(lr){9-10} \cmidrule(lr){11-12} \cmidrule(lr){13-14} 
& & F1\,$\uparrow$ & FPR\,$\downarrow$ & F1\,$\uparrow$ & FPR\,$\downarrow$ & F1\,$\uparrow$ & FPR\,$\downarrow$ & F1\,$\uparrow$ & FPR\,$\downarrow$ & F1\,$\uparrow$ & FPR\,$\downarrow$ & F1\,$\uparrow$ & FPR\,$\downarrow$  \\
\midrule

\multirow{2}{*}{\textbf{LSTM}} 
& NoAug & \textbf{0.9982} & \textbf{0.0002} & \textbf{0.9988} & \textbf{0.0003} & \textbf{0.9980} & \textbf{0.0003} & \textbf{0.8378} & \textbf{0.0020} & \textbf{0.9214} & 0.0792 & 0.9010 & 0.0690 \\
& Ours & 0.9872 & 0.0037 & 0.9965 & 0.0017 & 0.9791 & 0.0037 & 0.5733 & 0.0105 & 0.9211 & \textbf{0.0607} & \textbf{0.9088} & \textbf{0.0556} \\
\midrule

\multirow{2}{*}{\textbf{1D-CNN}} 
& NoAug & \textbf{0.9829} & \textbf{0.0029} & \textbf{0.9849} & \textbf{0.0054} & \textbf{0.9567} & \textbf{0.0048} & \textbf{0.4779} & 0.0020 & \textbf{0.7115} & \textbf{0.1678} & \textbf{0.7370} & \textbf{0.1389} \\
& Ours & 0.9297 & 0.0238 & 0.9509 & 0.0341 & 0.8623 & 0.0223 & 0.3093 & \textbf{0.0004} & 0.6983 & 0.2467 & 0.7069 & 0.1955 \\
\midrule

\multirow{2}{*}{\textbf{Transformer}} 
& NoAug & \textbf{0.9932} & \textbf{0.0013} & \textbf{0.9946} & \textbf{0.0014} & \textbf{0.9911} & \textbf{0.0017} & \textbf{0.7075} & 0.0056 & 0.8670 & \textbf{0.1092} & 0.8325 & \textbf{0.0779}  \\
& Ours & 0.9828 & 0.0053 & 0.9920 & 0.0055 & 0.9879 & 0.0027 & 0.6929 & \textbf{0.0008} & \textbf{0.8923} & 0.1108 & \textbf{0.8590} & 0.0954 \\
\midrule

\multirow{2}{*}{\textbf{DF}} 
& NoAug & \textbf{0.9968} & \textbf{0.0006} & \textbf{0.9979} & \textbf{0.0008} & \textbf{0.9964} & \textbf{0.0004} & \textbf{0.8058} & \textbf{0.0008} & 0.8807 & \textbf{0.1015} & 0.8843 & 0.0583  \\
& Ours & 0.9898 & 0.0022 & 0.9899 & 0.0054 & 0.9888 & 0.0006 & 0.6923 & 0.0116 & \textbf{0.8862} & 0.1070 & \textbf{0.9000} & \textbf{0.0578}  \\
\midrule

\multirow{2}{*}{\textbf{BAPM}} 
& NoAug & \textbf{0.9946} & \textbf{0.0009} & \textbf{0.9961} & \textbf{0.0020} & \textbf{0.9937} & \textbf{0.0011} & \textbf{0.7595} & 0.0068 & \textbf{0.8507} & 0.1462 & \textbf{0.8487} & 0.0826  \\
& Ours & 0.9606 & 0.0141 & 0.9580 & 0.0312 & 0.9206 & 0.0268 & 0.6286 & \textbf{0.0060} & 0.8503 & \textbf{0.1238} & 0.8481 & \textbf{0.0803}  \\
\midrule

\multirow{2}{*}{\textbf{TMWF}} 
& NoAug & \textbf{0.9962} & \textbf{0.0011} & \textbf{0.9982} & \textbf{0.0004} & \textbf{0.9957} & \textbf{0.0005} & \textbf{0.8050} & 0.0056 & \textbf{0.8926} & 0.1072 & 0.8857 & 0.0630  \\
& Ours & 0.9779 & 0.0073 & 0.9942 & 0.0019 & 0.9785 & 0.0016 & 0.6222 & \textbf{0.0048} & 0.8909 & \textbf{0.0961} & \textbf{0.8937} & \textbf{0.0604}  \\
\midrule

\multirow{2}{*}{\textbf{LUCID}} 
& NoAug & 0.8871 & 0.0048 & \textbf{0.9465} & 0.0150 & 0.9548 & 0.0123 & 0 & -- & 0 & -- & \textbf{0.4803} & \textbf{0.0336}  \\
& Ours & \textbf{0.8873} & \textbf{0.0041} & 0.9047 & \textbf{0.0097} & \textbf{0.9610} & \textbf{0.0066} & 0 & -- & 0 & -- & 0.4683 & 0.0337 \\
\midrule

\multirow{2}{*}{\textbf{Whisper}} 
& NoAug & \textbf{0.9902} & \textbf{0.0006} & \textbf{0.9924} & \textbf{0.0018} & \textbf{0.9939} & \textbf{0.0019} & 0 & -- & \textbf{0.8469} & 0.1794 & 0.1521 & 0.0360  \\
& Ours & 0.9638 & 0.0046 & 0.9715 & 0.0081 & 0.9807 & 0.0133 & 0 & -- & 0.8459 & \textbf{0.1584} & \textbf{0.7369} & \textbf{0.0315}  \\
\midrule

\multirow{2}{*}{\textbf{MATEC}} 
& NoAug & 0.9933 & 0.0003 & 0.9952 & 0.0006 & 0.9965 & 0.0006 & 0.7692 & \textbf{0.0028} & 0.8861 & 0.1237 & 0.8566 & 0.0949  \\
& Ours & \textbf{0.9967} & \textbf{0.0001} & \textbf{0.9971} & \textbf{0.0004} & \textbf{0.9976} & \textbf{0.0005} & \textbf{0.8267} & \textbf{0.0028} & \textbf{0.8963} & \textbf{0.0995} & \textbf{0.8782} & \textbf{0.0836}  \\
\midrule

\multirow{2}{*}{\textbf{SmartDetector}} 
& NoAug & 0.9990 & 0.0001 & \textbf{0.9999} & \textbf{0} & 0.9998 & 0.0001 & 0.7397 & 0.0044 & 0.8589 & \textbf{0.1276} & 0.8978 & 0.0321  \\
& Ours & \textbf{0.9999} & \textbf{0} & \textbf{0.9999} & 0.0001 & \textbf{0.9999} & \textbf{0} & \textbf{0.8112} & \textbf{0.0016} & \textbf{0.8850} & 0.1309 & \textbf{0.9186} & \textbf{0.0290}  \\

\bottomrule
\end{tabular}
}
\begin{tablenotes}
\small
\setlength{\leftskip}{4.1em}
\setlength{\rightskip}{1em}
\item * \textbf{Aug}: Augmentation. \textbf{NoAug}: No augmentation.
\item ** The best results are marked in \textbf{bold}. We mark the FPR as ``--'' when the F1 score is 0 (as FPR is meaningless in this case). 
\end{tablenotes}
\vspace{-2mm}
\end{table*}

\subsection{Evaluation in Real-World Environment} \label{sec:real-world experiment}
To better understand the practicality of our method, we evaluate it on two datasets ($\mathcal{D} _5$ and $\mathcal{D}_6$) collected in real-world network environments. Traffic collected in real-world environments may suffer from inaccurate labeling (especially on MAWILab, i.e., $\mathcal{D} _5$). Moreover, since the training and test sets are collected on different dates, the distribution shift is more pronounced. The evaluation is similar to that in Section \ref{sec:augment performance}, and the results are shown in Table \ref{tab:real-world compare}. More detailed metrics for our method can be found in Table \ref{tab:ood_results_full} in Appendix \ref{app:detailed results}.

Compared to the non-augmented baseline, our method enhances the F1 score in most cases without significantly increasing the FPR; in many instances, our method even reduces the FPR. Our method also demonstrates advantages over other data augmentation techniques. For the Whisper model on dataset $\mathcal{D}_6$, our approach increases the F1 score from 0.255 to 0.6338 and reduces the FPR from 0.1052 to 0.0516. In contrast, Mixup, the best-performing baseline, only improves the F1 score to 0.2579 and even increases the FPR. For the 1D-CNN model, most augmentations reduce the F1 score, which may indicate that 1D-CNN is not well adapted to the variations introduced by data augmentations and instead treats them as noise. LUCID fails on $\mathcal{D}_5$, possibly because some of the features it uses are not informative on this dataset (e.g., relative time from the first packet), thereby impairing its detection capability. Whisper leverages frequency-domain features for detection, whereas FreqMix performs poorly on Whisper, yielding an FPR of 1 on $\mathcal{D}_5$ and failing on $\mathcal{D}_6$. This suggests that FreqMix's mixing strategy may be overly simplistic, thereby disrupting Whisper's ability to capture informative frequency-domain features. 
Overall, our approach still shows some effectiveness on real-world datasets and improves the generalization of detection models in practical environments.

\rev{Similar to Section \ref{sec:augment performance}, we plot the ROC curves of several representative models to investigate the impact of FPR, as shown in the bottom row of Figure \ref{fig:roc_D1_D6}. We choose dataset $\mathcal{D}_6$ because, on $\mathcal{D}_5$, our method improves FPR in most cases. As shown in Figure \ref{fig:roc_D1_D6}, the ROC curves on the real-world dataset are generally closer to each other, especially at low FPR levels, indicating that the performance differences among these methods are relatively small. Nevertheless, our method still shows some advantage on SmartDetector. Our method also consistently achieves slightly better performance on LSTM. Unlike the artificial dataset, our method can outperform the model trained without augmentation even at very low FPR levels on the real-world dataset.}

\rev{The analysis on both the artificial and real-world datasets in Figure \ref{fig:roc_D1_D6} indicates that it is feasible to constrain the FPR of our method to a desired level by adjusting the decision threshold. This provides a way to balance FPR and TPR for our method. Under the same FPR level, our method outperforms the model without augmentation in most cases, thereby still remaining effective in improving the model's performance. In practical applications, we recommend evaluating our method on a validation set according to the target requirements and adjusting the decision threshold to achieve an acceptable FPR-TPR trade-off for better deployment. }

\subsection{Ablation Study}

To verify the contribution of each component in our method, we conduct an ablation study. To eliminate the potential interference from data scarcity and label noise, we perform experiments on dataset $\mathcal{D}_1\sim\mathcal{D}_3$. The experimental results are presented in Figure \ref{fig:ablation_study}. ``Full Setting'' refers to our complete data augmentation method. ``w/o Linear Mixing'' means that when the selected two samples have different labels, frequency-domain mixing is used instead, while the other components remain the same as in the full setting. ``w/o Pre-Aug'' removes the pre-augmentation component from the full setting. ``w/o Windowing'' keeps the other components the same but does not perform windowing.

Figure \ref{fig:ablation_study} shows that the contribution of each component varies under different conditions. In most cases, the full setting achieves the best performance. This indicates that the combination of pre-augmentation, inter-class linear mixing, and windowing works synergistically to improve model performance. Compared to other settings, the full setting may lead to an increased FPR, as it enhances the model's ability to identify malicious samples to a greater extent, thereby introducing some additional false positives. However, in many cases, the FPR under the full setting is not the highest. This suggests that the combination of these operations is in fact beneficial for suppressing the rise in FPR.

Removing linear mixing typically leads to a substantial drop in F1, highlighting the importance of maintaining inter-class linear mixing and thereby validating limitation (2) proposed in Section \ref{sec:limitations}. Removing pre-augmentation and windowing has a relatively smaller impact on the F1 score, but can still degrade the performance in most cases. In certain scenarios, such as MATEC on dataset $\mathcal{D}_1$, they are critical components. Both pre-augmentation and windowing contribute to improving the stability of our method. Furthermore, windowing appears to have a significant effect on the FPR; in many cases, its removal results in a sharp increase in FPR. In our augmentation, windowing enables the model to better focus on the local behaviors of packet feature sequences, which may allow more precise identification of malicious traffic and thus reduce the FPR.

\subsection{Impact on In-Distribution Generalization} \label{sec:impact on in-distribution}

Our data augmentation method introduces certain variations to the samples, which facilitates the learning of more robust features. However, Tsipras et al. \cite{conf/iclr/TsiprasSETM19} reported that leveraging more robust features does not necessarily improve the model’s accuracy on standard data (i.e., in-distribution data in our experiments). In some cases, especially in adversarial training, there can be a trade-off between OOD generalization and ID generalization. To investigate the impact of our augmentation on the model’s ID generalization, we evaluate the performance of each model on ID data before and after augmentation. The results are presented in Table \ref{tab:augmentation_in_distribution}. More detailed metrics can be found in Table \ref{tab:id_results_full} in Appendix \ref{app:detailed results}.

As shown in Table \ref{tab:augmentation_in_distribution}, models generally achieve good performance on ID data without data augmentation. 
By comparing the data across Table \ref{tab:method_compare}, Table \ref{tab:real-world compare},  and Table \ref{tab:augmentation_in_distribution}, we observe that although the models perform well on ID data without augmentation, their performance decreases significantly when they face a distribution shift. 
Table \ref{tab:augmentation_in_distribution} also shows that while our data augmentation improves OOD generalization, it may lead to a decrease in performance on ID data. However, this degradation is often minor: in many cases, the F1 score decreases by less than 3\%, and the FPR increases by less than 0.5\%. Moreover, our method can actually improve ID performance in a number of cases. Notably, the F1 score variations on $\mathcal{D}_4$ are more pronounced compared to other datasets, which can be attributed to its relatively small test set size (only 81 malicious samples). 
The phenomenon of performance degradation on ID data closely resembles the trade-off observed in adversarial training, where improved adversarial accuracy (OOD performance in our situation) comes at the cost of reduced standard accuracy (ID performance in our situation) \cite{conf/iclr/TsiprasSETM19,conf/icml/RiceWK20,conf/eccv/SuZCYCG18}. Generally, adversarial samples can be regarded as worst-case OOD data \cite{GeirhosJMZBBW20}. In our experiments, this phenomenon suggests that some augmented samples might act like adversarial samples. Drawing a parallel to adversarial training, and considering that our method significantly improves OOD generalization while having only a minor impact on ID performance, we argue that this trade-off is worthwhile, as it enhances the model’s practical applicability in real-world scenarios.

\section{Discussion}


\noindent \textbf{Biases of data augmentation.} Our augmentation may generate samples that would not occur in reality, potentially introducing some biases to the actual data. However, during frequency-domain mixing between intra-class samples, our goal is to encourage the model to focus on low-frequency features. Although some generated samples may not exist in reality, they preserve the relative stability of the low-frequency components and still help the model learn robust low-frequency representations. As for the linear mixing between inter-class samples, this is essentially the idea of Mixup, which has proven effective in prior studies \cite{ZhangDK022,mixuptraining}.

\noindent \textbf{Applicability of our method.} In principle, our method is applicable to most numerical sequence features, such as packet length sequences and IAT sequences. Categorical features, such as protocol types, are not suitable for our approach. A potential concern is whether the augmentation might disrupt the correlations among features when a model utilizes multiple features simultaneously. Our intuition is that, for a given packet, the correlations between different features are usually weak. For example, the protocol may have some influence on packet length, but it is far from being determinative. Moreover, our experiments in Section \ref{sec:experiments} demonstrate that augmenting only the applicable features remains effective. Exploring the impacts of cross-feature correlations is a possible direction for future work.

\noindent \textbf{Robustness against evasion attacks.} Attackers may employ various techniques to evade detection. As defending against such attacks is beyond the primary scope of this paper, we have not evaluated our method in specific evasion scenarios. However, given the parallels between our approach and adversarial training discussed in Section \ref{sec:impact on in-distribution}, we believe that our method may offer a certain degree of robustness against evasion attacks, much like adversarial training can defend against adversarial examples. This is also a potential direction for future work.

\noindent \textbf{Limitations of our method.} First, our method has limited effectiveness on very short packet sequences. This is because shorter sequences provide fewer frequency components for mixing, which constrains the diversity of the augmented samples. We apply windowing to reduce the requirement for sequence length and recommend using sequences consisting of at least six packets. Furthermore, our method aims to enhance the model’s ability to recognize the OOD flows described in Section \ref{sec:ideas to solve}, and may be ineffective against severely drifted flows that differ substantially (particularly in their low-frequency components) from the training traffic.

\section{Conclusion} \label{sec:conclusion}
In this paper, we introduce the frequency-domain mixing data augmentation method into malicious traffic detection to improve the out-of-distribution generalization ability of detection models. We first provide a theoretical analysis of this approach and identify its limitations in supporting network traffic classification. Based on this analysis, we propose an improved augmentation method for network packet feature sequences. Our method incorporates a pre-augmentation step to increase data diversity, and adopts a new mixing strategy in which both samples and labels are linearly mixed for inter-class samples, whereas frequency-domain mixing is applied for intra-class samples. In addition, we employ a windowed Fourier transform for frequency-domain conversion to accommodate the nonstationarity of network traffic sequences. We conduct extensive experiments on multiple artificial and real-world datasets, and the results demonstrate that our method can significantly improve the out-of-distribution generalization ability of malicious traffic detection models and outperforms other data augmentation methods. The improvement in the out-of-distribution generalization ability enables the model to better handle the complex network traffic in real-world environments. In future work, we plan to explore better frequency-domain mixing schemes for network traffic data augmentation.



\bibliographystyle{ACM-Reference-Format}
\bibliography{ref}

\appendix 

\section{Open Science} 
Our code is available at \url{https://github.com/Yebmoon/Freqmix_improved}. 

\section{Ethical Considerations} 

Our data collection and experimental procedures were conducted with authorization and caused no harm to real-world systems. Therefore, this study raises no ethical concerns.




\section{Experimental Details} \label{app:experimental details}
For all the divided datasets, we first extract the packet feature sequence of each session. Following the general practice in traffic classification \cite{XieCDX00SZ23}, we uniformly set the length of the extracted packet feature sequences to 100. Sequences shorter than 100 are padded with zeros at the end, while those longer than 100 are truncated. In our experiments, the specific pre-augmentation strategies we adopt are as follows:
\begin{itemize}[left=0pt]
\item Packet shift: Randomly select a certain number of positions (determined by the pre-augmentation ratio) and shift the features at these positions forward or backward by at most 2 positions.
\item Packet variation: Randomly select a certain number of positions (determined by the pre-augmentation ratio), compute the standard deviation $std$ of the sequence, and generate a random value $r$ following a normal distribution $\mathcal{N}(0,1)$ for each selected position. With a scaling factor of 0.02, we add the final noise, calculated as $std \cdot r \cdot 0.02$,  to the corresponding positions.
\end{itemize}
These parameter settings for pre-augmentation are chosen to constrain the augmentation strength within a relatively small range, ensuring that the perturbations do not excessively change the sequence and thus do not violate label consistency. One can adjust the pre-augmentation algorithms to suit different scenarios.

\begin{table}[htbp]
\centering
\caption{Parameter settings for data augmentation methods.}
\vspace{-1.5mm}
\label{tab:Parameter Settings}
\begin{tabular}{cl}
\toprule
\textbf{Method} & \textbf{Parameter Settings} \\
\midrule
Our Method & \makecell[l]{Pre-augmentation ratio: 0.2 \\ Window length $w$: 10 \\ Mixing ratio: beta distribution \\ with parameters (1, 1)} \\
\midrule
Jittering & \makecell[l]{Noise distribution: Gaussian \\ distribution $\mathcal{N}(0,1)$ and \\ scaled by  0.05 times the sequence \\ standard deviation} \\
\midrule
Permutation & Window length: 5 \\
\midrule
Random Masking & Ratio of masked elements: 0.2 \\
\midrule
Mixup & \makecell[l]{Mixing ratio: beta distribution \\ with parameters (1, 1)} \\
\midrule
\makecell[c]{FreqMix} & \makecell[l]{Mixing ratio: beta distribution \\ with parameters (1, 1)} \\
\midrule
Rosetta & Same as \cite{XieCDX00SZ23} \\
\bottomrule
\end{tabular}
\vspace{-2mm}
\end{table}

\begin{figure}[b]
    \centering
    \includegraphics[width=0.8\linewidth]{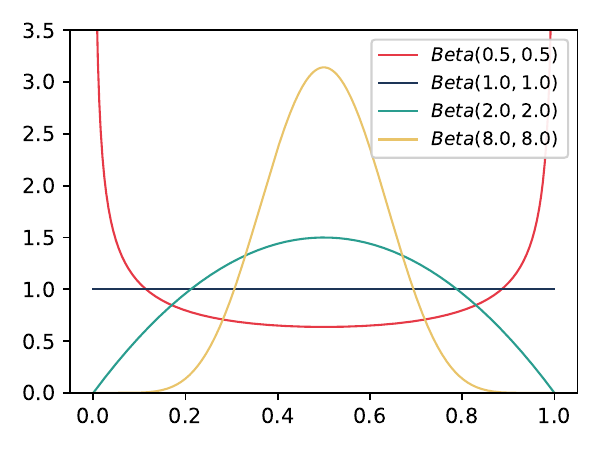}
    \vspace{-2mm}
    \caption{Probability density curves of beta distributions with different parameters.}
    \vspace{-2mm}
    \label{fig:beta_distribution}
\end{figure}
\begin{figure*}[hbtp]
    \centering
    \includegraphics[width=\linewidth]{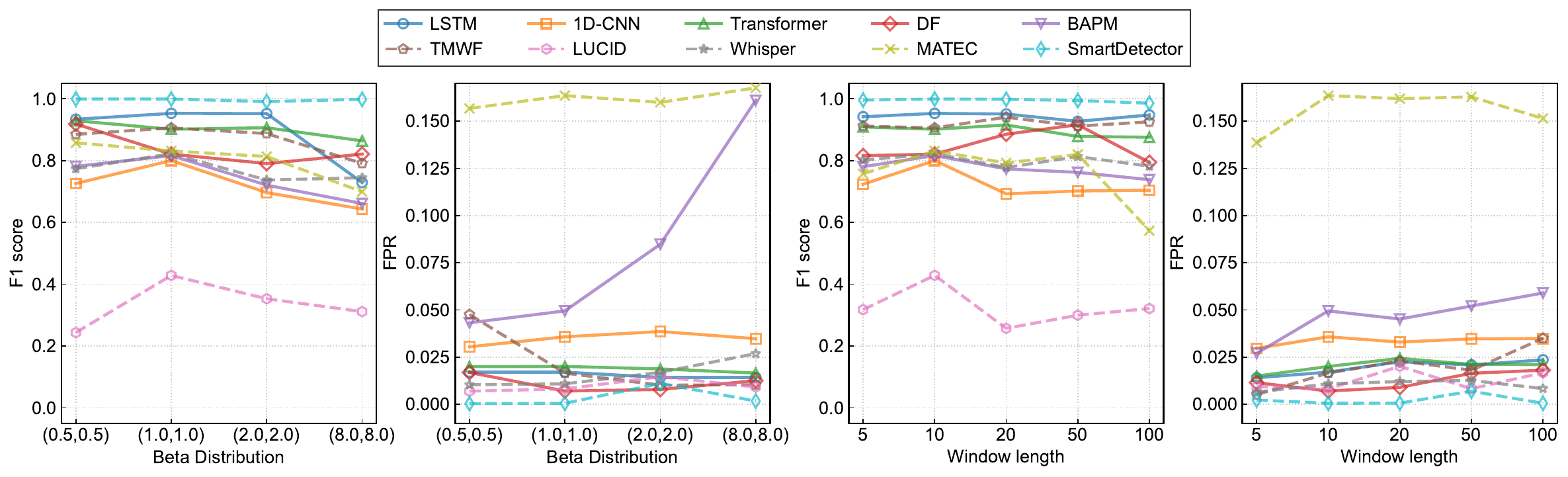}
    \vspace{-3mm}
    \caption{Impact of different parameter values on model performance.}
    \label{fig:sensitivity analysis}
    \vspace{-3mm}
\end{figure*}

When training the models, we set the learning rate to 0.0001 and use the Adam optimizer for optimization. We adopt an early stopping strategy, where training is terminated if the validation loss does not decrease in 20 epochs. For the seven specialized models (DF, BAPM, TMWF, LUCID, Whisper, MATEC, and SmartDetector), all hyperparameters remain the same as those in their original papers (note that we use the supervised version implemented by Yan et al. \cite{CertTA} for Whisper), except for adjustments necessitated by the input sequence length being set to 100. The training of SmartDetector consists of a contrastive learning stage and a supervised learning stage. We apply data augmentation during the contrastive learning stage, while no augmentation is performed during the supervised stage. The key parameter settings for each data augmentation method are listed in Table \ref{tab:Parameter Settings}. Our method has two main parameters in addition to the pre-augmentation ratio (0.2 means that 20\% of the sequence elements are processed with pre-augmentation operations): the Fourier transform window length $w$ and the frequency-domain mixing ratio $\lambda$. Their roles are detailed in Algorithm \ref{algorithm1}. Since the mixing ratio in Mixup \cite{ZhangCDL18} is drawn from a beta distribution, we also apply a beta distribution to the mixing ratio in our method.

\section{Detailed Results of Our Method's Augmentation Performance on Out-of-Distribution and In-Distribution Data} \label{app:detailed results}

The detailed results of model performance on out-of-distribution (Section \ref{sec:augment performance} and Section \ref{sec:real-world experiment}) and in-distribution (Section \ref{sec:impact on in-distribution}) data are presented in Table \ref{tab:ood_results_full}  and Table \ref{tab:id_results_full}, respectively.

Table \ref{tab:ood_results_full} shows that after applying our data augmentation method, models exhibit a considerable improvement when handling out-of-distribution samples. In most cases, our method substantially enhances the model's recall while preserving precision, which ultimately leads to an increase in F1 score. This suggests that the model's ability to identify malicious samples has been strengthened. After augmentation, the FPR generally shows a slight increase, which is a side effect of the improved capability to identify malicious samples. Overall, our data augmentation maintains a relatively low false positive rate while significantly improving the recall.

Table \ref{tab:id_results_full} shows that in most cases, applying our data augmentation method leads to a decrease in recall and precision on in-distribution data, resulting in a drop in F1 score, and the FPR also increases. However, both the decrease and the increase are usually slight. Moreover, there are several cases where our method actually improves performance on in-distribution data. Since the augmentation significantly improves the model's out-of-distribution generalization ability, we argue that this trade-off is worthwhile.
\section{Parameter Sensitivity Analysis} \label{app:parameter sensitivity}

    
To understand the impact of parameters on our method, we conduct a parameter sensitivity analysis on dataset $\mathcal{D}_1$. Two primary parameters influence our method: the frequency-domain mixing ratio $\lambda$ and the Fourier transform window length $w$ (see Algorithm \ref{algorithm1}). 

Following the practice of Mixup, the mixing ratio $\lambda$ is sampled from a beta distribution, $\operatorname{Beta}(\alpha,\beta)$, meaning that the mixing ratio is governed by the parameters of this distribution. For the beta distribution $\operatorname{Beta}(\alpha,\beta)$, $\alpha$ and $\beta$ are shape parameters that primarily control the shape of the probability density function near $x = 0$ and $x = 1$, respectively. Considering the symmetry of mixing, we focus on beta distributions where $\alpha$ and $\beta$ are equal. We conduct experiments with $\lambda$ sampled from beta distributions with parameters 0.5, 1.0, 2.0, and 8.0. The probability density functions of these beta distributions are visualized in Figure \ref{fig:beta_distribution}. For the distribution $\lambda\sim \operatorname{Beta}(\alpha,\alpha)$, when $\alpha=1$, it is a uniform distribution; when $\alpha= 0.5$ , $\lambda$ is more likely to be sampled near the endpoints, resulting in a smaller mixing ratio, and the generated data will be closer to the original samples; when $\alpha=2.0$, $\lambda$ is more likely to be sampled near the center, leading to a larger mixing ratio, and the generated data will be quite different from the original samples; when $\alpha=8.0$, compared with $\alpha=2.0$, the tendency to sample near the center becomes even more pronounced. A small $\alpha$ may generate samples that are easier for the model to learn but might lack sufficient diversity. Conversely, a large $\alpha$ may lead to higher diversity but risks producing implausible samples. The experimental results are shown in Figure \ref{fig:sensitivity analysis}. For the window length $w$, we experiment with values of 5, 10, 20, 50, and 100 (where $w=100$ is equivalent to no windowing), and the results are also shown in Figure \ref{fig:sensitivity analysis}.

Figure \ref{fig:sensitivity analysis} indicates that model performance is relatively stable under moderate parameter settings, while more aggressive settings, such as 
$\operatorname{Beta}(8.0,8.0)$ or a window length of 100, lead to more noticeable performance fluctuations. There are also cases where a model performs significantly better under a specific parameter setting than under others, likely because the augmented data generated under that setting is more similar to out-of-distribution samples and can be better utilized by the model.

\begin{table*}[htbp]
\centering
\caption{Model performance on out-of-distribution data before and after applying our proposed data augmentation method.}
\label{tab:ood_results_full}
\vspace{-1.5mm}
\setlength{\tabcolsep}{2pt}
\renewcommand{\arraystretch}{0.75}
\resizebox{\textwidth}{!}{
\begin{tabular}{ll cccccccc cccccccc}
\toprule
\multirow{2}{*}{\textbf{Model}} & \multirow{2}{*}{\textbf{Metric}} & \multicolumn{2}{c}{$\mathcal{D}_1$} & \multicolumn{2}{c}{$\mathcal{D}_2$} & \multicolumn{2}{c}{$\mathcal{D}_3$} & \multicolumn{2}{c}{$\mathcal{D}_4$} & \multicolumn{2}{c}{$\mathcal{D}_5$} & \multicolumn{2}{c}{$\mathcal{D}_6$} \\
\cmidrule(lr){3-4} \cmidrule(lr){5-6} \cmidrule(lr){7-8} \cmidrule(lr){9-10} \cmidrule(lr){11-12} \cmidrule(lr){13-14}
& & NoAug & Ours & NoAug & Ours & NoAug & Ours & NoAug & Ours & NoAug & Ours & NoAug & Ours \\
\midrule
\multirow{4}{*}{\textbf{LSTM}} & Recall & 0.5950 & \textbf{0.9678} & 0.2406 & \textbf{0.8789} & 0.4081 & \textbf{0.6987} & 0 & 0 & \textbf{0.8566} & 0.8297 & 0.7148 & \textbf{0.7149} \\
& Precision & \textbf{0.9889} & 0.9382 & \textbf{0.9814} & 0.9611 & \textbf{0.9990} & 0.9776 & 0 & 0 & 0.6923 & \textbf{0.7142} & 0.7887 & \textbf{0.8166} \\
& F1 & 0.7429 & \textbf{0.9528} & 0.3865 & \textbf{0.9182} & 0.5794 & \textbf{0.8149} & 0 & 0 & 0.7657 & \textbf{0.7677} & 0.7499 & \textbf{0.7623} \\
& FPR & \textbf{0.0018} & 0.0171 & \textbf{0.0006} & 0.0045 & \textbf{0.0001} & 0.0035 & -- & -- & 0.1831 & \textbf{0.1597} & 0.1574 & \textbf{0.1320} \\ \midrule
\multirow{4}{*}{\textbf{1D-CNN}} & Recall & 0.5904 & \textbf{0.7558} & 0.4548 & \textbf{0.7032} & 0.7082 & \textbf{0.9683} & 0.0060 & \textbf{0.4337} & 0.5448 & \textbf{0.5990} & 0.5891 & \textbf{0.6291} \\
& Precision & \textbf{0.9658} & 0.8499 & \textbf{0.9022} & 0.6101 & \textbf{0.9744} & 0.8947 & 0.0556 & \textbf{0.8372} & \textbf{0.5978} & 0.5331 & \textbf{0.6635} & 0.6110 \\
& F1 & 0.7328 & \textbf{0.8001} & 0.6047 & \textbf{0.6533} & 0.8202 & \textbf{0.9301} & 0.0109 & \textbf{0.5714} & \textbf{0.5701} & 0.5641 & \textbf{0.6240} & 0.6199 \\
& FPR & \textbf{0.0056} & 0.0358 & \textbf{0.0063} & 0.0571 & \textbf{0.0041} & 0.0253 & 0.0013 & \textbf{0.0011} & \textbf{0.1763} & 0.2523 & \textbf{0.2457} & 0.3293 \\ \midrule
\multirow{4}{*}{\textbf{Transformer}} & Recall & 0.6011 & \textbf{0.8821} & 0.4102 & \textbf{0.9159} & 0.5485 & \textbf{0.6479} & 0 & \textbf{0.4337} & 0.7517 & \textbf{0.8231} & 0.6125 & \textbf{0.6730} \\
& Precision & \textbf{0.9643} & 0.9215 & \textbf{0.9298} & 0.9129 & \textbf{0.9900} & 0.9856 & 0 & \textbf{0.3380} & \textbf{0.6520} & 0.6510 & \textbf{0.7591} & 0.7353 \\
& F1 & 0.7405 & \textbf{0.9014} & 0.5692 & \textbf{0.9144} & 0.7059 & \textbf{0.7818} & 0 & \textbf{0.3799} & 0.6983 & \textbf{0.7270} & 0.6780 & \textbf{0.7027} \\
& FPR & \textbf{0.0060} & 0.0201 & \textbf{0.0039} & 0.0111 & \textbf{0.0012} & 0.0021 & -- & \textbf{0.0108} & \textbf{0.1929} & 0.2122 & \textbf{0.1599} & 0.1992 \\ \midrule
\multirow{4}{*}{\textbf{DF}} & Recall & 0.5798 & \textbf{0.7148} & 0.5928 & \textbf{0.7770} & 0.4689 & \textbf{0.5674} & 0 & \textbf{0.3675} & 0.7532 & \textbf{0.7967} & \textbf{0.7031} & 0.7014 \\
& Precision & 0.9346 & \textbf{0.9644} & \textbf{0.9643} & 0.9252 & \textbf{0.9972} & 0.9947 & 0 & \textbf{0.0483} & 0.6649 & \textbf{0.6886} & 0.8034 & \textbf{0.8111} \\
& F1 & 0.7156 & \textbf{0.8210} & 0.7342 & \textbf{0.8446} & 0.6379 & \textbf{0.7226} & 0 & \textbf{0.0854} & 0.7063 & \textbf{0.7387} & 0.7499 & \textbf{0.7523} \\
& FPR & 0.0109 & \textbf{0.0071} & \textbf{0.0028} & 0.0080 & \textbf{0.0003} & 0.0007 & -- & \textbf{0.0919} & 0.1825 & \textbf{0.1733} & 0.1415 & \textbf{0.1343} \\ \midrule
\multirow{4}{*}{\textbf{BAPM}} & Recall & 0.5623 & \textbf{0.8167} & 0.4233 & \textbf{0.7693} & 0.5130 & \textbf{0.7571} & 0.3675 & \textbf{0.8012} & \textbf{0.7626} & 0.7506 & 0.6768 & \textbf{0.7016} \\
& Precision & \textbf{0.9475} & 0.8158 & \textbf{0.9128} & 0.6513 & \textbf{0.9927} & 0.8759 & 0.0857 & \textbf{0.2519} & 0.6050 & \textbf{0.6352} & \textbf{0.7701} & 0.7575 \\
& F1 & 0.7058 & \textbf{0.8163} & 0.5783 & \textbf{0.7054} & 0.6764 & \textbf{0.8121} & 0.1390 & \textbf{0.3833} & 0.6747 & \textbf{0.6881} & 0.7204 & \textbf{0.7285} \\
& FPR & \textbf{0.0084} & 0.0495 & \textbf{0.0051} & 0.0523 & \textbf{0.0008} & 0.0238 & 0.0498 & \textbf{0.0302} & 0.2394 & \textbf{0.2073} & \textbf{0.1662} & 0.1846 \\ \midrule
\multirow{4}{*}{\textbf{TMWF}} & Recall & 0.7558 & \textbf{0.8793} & 0.6239 & \textbf{0.7394} & 0.5176 & \textbf{0.5990} & 0 & \textbf{0.8072} & 0.7964 & \textbf{0.7991} & 0.6683 & \textbf{0.6871} \\
& Precision & 0.9261 & \textbf{0.9341} & \textbf{0.9778} & 0.9634 & \textbf{0.9972} & 0.9898 & 0 & \textbf{0.3583} & 0.6725 & \textbf{0.6752} & 0.7985 & \textbf{0.8001} \\
& F1 & 0.8323 & \textbf{0.9059} & 0.7617 & \textbf{0.8366} & 0.6815 & \textbf{0.7463} & 0 & \textbf{0.4963} & 0.7292 & \textbf{0.7320} & 0.7276 & \textbf{0.7393} \\
& FPR & \textbf{0.0162} & 0.0166 & \textbf{0.0018} & 0.0036 & \textbf{0.0003} & 0.0014 & -- & \textbf{0.0183} & 0.1866 & \textbf{0.1848} & \textbf{0.1387} & 0.1412 \\ \midrule

\multirow{4}{*}{\textbf{LUCID}} 
& Recall    & 0.1975 & \textbf{0.2764} & 0.1918 & \textbf{0.2546} & 0.6951 & \textbf{0.7299} & 0 & 0 & 0 & 0 & \textbf{0.2414} & 0.2219 \\
& Precision & 0.9110 & \textbf{0.9531} & 0.8133 & \textbf{0.9033} & 0.9138 & \textbf{0.9471} & 0 & 0 & 0 & 0 & 0.7021 & \textbf{0.7067} \\
& F1        & 0.3247 & \textbf{0.4285} & 0.3104 & \textbf{0.3973} & 0.7896 & \textbf{0.8245} & 0 & 0 & 0 & 0 & \textbf{0.3593} & 0.3378 \\
& FPR       & 0.0116 & \textbf{0.0082} & 0.0150 & \textbf{0.0093} & 0.0115 & \textbf{0.0071} & -- & -- & -- & -- & 0.0496 & \textbf{0.0446} \\
\midrule

\multirow{4}{*}{\textbf{Whisper}} 
& Recall    & 0.5678 & \textbf{0.7104} & 0.6121 & \textbf{0.6784} & 0.7105 & \textbf{0.9659} & 0 & 0 & 0.6061 & \textbf{0.6403} & 0.1762 & \textbf{0.5108} \\
& Precision & \textbf{0.9879} & 0.9758 & \textbf{0.9859} & 0.9655 & \textbf{0.9618} & 0.8561 & 0 & 0 & 0.3350 & \textbf{0.3904} & 0.4610 & \textbf{0.8349} \\
& F1        & 0.7211 & \textbf{0.8222} & 0.7553 & \textbf{0.7969} & 0.8173 & \textbf{0.9077} & 0 & 0 & 0.4315 & \textbf{0.4850} & 0.2550 & \textbf{0.6338} \\
& FPR       & \textbf{0.0043} & 0.0109 & \textbf{0.0031} & 0.0086 & \textbf{0.0042} & 0.0240 & -- & -- & 0.2024 & \textbf{0.1682} & 0.1052 & \textbf{0.0516} \\
\midrule

\multirow{4}{*}{\textbf{MATEC}} 
& Recall    & 0.2059 & \textbf{0.8987} & 0.0796 & \textbf{0.9548} & 0.8076 & \textbf{0.8861} & 0.2169 & \textbf{0.4337} & 0.8076 & \textbf{0.8083} & 0.6604 & \textbf{0.6639} \\
& Precision & 0.5354 & \textbf{0.7731} & 0.9490 & \textbf{0.9958} & 0.9933 & \textbf{0.9961} & \textbf{0.7826} & 0.6486 & 0.6215 & \textbf{0.6567} & \textbf{0.7105} & 0.7090 \\
& F1        & 0.2974 & \textbf{0.8312} & 0.1469 & \textbf{0.9748} & 0.8909 & \textbf{0.9379} & 0.3396 & \textbf{0.5199} & 0.7024 & \textbf{0.7247} & 0.6845 & \textbf{0.6857} \\
& FPR       & \textbf{0.1108} & 0.1635 & 0.0015 & \textbf{0.0014} & 0.0008 & \textbf{0.0005} & \textbf{0.0008} & 0.0030 & 0.2365 & \textbf{0.2032} & \textbf{0.2213} & 0.2241 \\
\midrule

\multirow{4}{*}{\textbf{SmartDetector}} 
& Recall    & 0.9897 & \textbf{0.9994} & 0.9955 & \textbf{0.9999} & 0.9998 & \textbf{0.9999} & 0 & \textbf{0.4337} & 0.6217 & \textbf{0.6951} & 0.7425 & \textbf{0.8225} \\
& Precision & 0.9973 & \textbf{0.9990} & \textbf{1} & \textbf{1} & 0.9989 & \textbf{0.9992} & 0 & \textbf{0.4186} & 0.2811 & \textbf{0.3630} & 0.8072 & \textbf{0.8090} \\
& F1        & 0.9935 & \textbf{0.9992} & 0.9977 & \textbf{0.9999} & 0.9994 & \textbf{0.9996} & 0 & \textbf{0.4260} & 0.3871 & \textbf{0.4769} & 0.7735 & \textbf{0.8157} \\
& FPR       & 0.0017 & \textbf{0.0006} & \textbf{0} & \textbf{0} & 0.0002 & \textbf{0.0001} & -- & \textbf{0.0076} & 0.2675 & \textbf{0.2052} & \textbf{0.0906} & 0.0991 \\
\bottomrule
\end{tabular}
}
\begin{tablenotes}
\setlength{\leftskip}{-1em}
\setlength{\rightskip}{1em}
\small
\item * \textbf{NoAug}: No augmentation. 
\item ** The best results are marked in \textbf{bold}. We mark the FPR as ``--'' when the F1 score is 0 (as FPR is meaningless in this case). 
\end{tablenotes}
\end{table*}

For the beta distribution $\operatorname{Beta}(\alpha,\alpha)$, as $\alpha$ increases beyond 1.0 (uniform distribution), the generated samples deviate further from the original data. Consequently, several models such as LSTM, 1D-CNN, and LUCID experience substantial drops in F1 score. The impact of $\alpha$ on FPR is generally less pronounced, although models like BAPM and TMWF remain sensitive. The FPR of BAPM increases rapidly for larger $\alpha$, while TMWF exhibits relatively high FPR at $\alpha=0.5$. Overall, the uniform distribution ($\alpha=1.0$) provides a good balance between diversity and plausibility of the generated data, and is therefore recommended in our method.

\begin{table*}[hbtp]
\centering
\caption{Model performance on in-distribution data before and after applying our proposed data augmentation method.}
\label{tab:id_results_full}
\vspace{-1.5mm}
\setlength{\tabcolsep}{2pt}
\renewcommand{\arraystretch}{0.75}
\resizebox{\textwidth}{!}{
\begin{tabular}{ll cccccccc cccccccc}
\toprule
\multirow{2}{*}{\textbf{Model}} & \multirow{2}{*}{\textbf{Metric}} & \multicolumn{2}{c}{$\mathcal{D}_1$} & \multicolumn{2}{c}{$\mathcal{D}_2$} & \multicolumn{2}{c}{$\mathcal{D}_3$} & \multicolumn{2}{c}{$\mathcal{D}_4$} & \multicolumn{2}{c}{$\mathcal{D}_5$} & \multicolumn{2}{c}{$\mathcal{D}_6$} \\
\cmidrule(lr){3-4} \cmidrule(lr){5-6} \cmidrule(lr){7-8} \cmidrule(lr){9-10} \cmidrule(lr){11-12} \cmidrule(lr){13-14}
& & NoAug & Ours & NoAug & Ours & NoAug & Ours & NoAug & Ours & NoAug & Ours & NoAug & Ours \\
\midrule
\multirow{4}{*}{\textbf{LSTM}} & Recall & \textbf{0.9975} & 0.9939 & \textbf{0.9984} & 0.9973 & \textbf{0.9978} & 0.9798 & \textbf{0.7654} & 0.5309 & \textbf{0.9213} & 0.9052 & \textbf{0.8985} & 0.8973 \\
& Precision & \textbf{0.9989} & 0.9806 & \textbf{0.9993} & 0.9956 & \textbf{0.9981} & 0.9784 & \textbf{0.9254} & 0.6232 & 0.9214 & \textbf{0.9376} & 0.9035 & \textbf{0.9207} \\
& F1 & \textbf{0.9982} & 0.9872 & \textbf{0.9988} & 0.9965 & \textbf{0.9980} & 0.9791 & \textbf{0.8378} & 0.5733 & \textbf{0.9214} & 0.9211 & 0.9010 & \textbf{0.9088} \\
& FPR & \textbf{0.0002} & 0.0037 & \textbf{0.0003} & 0.0017 & \textbf{0.0003} & 0.0037 & \textbf{0.0020} & 0.0105 & 0.0792 & \textbf{0.0607} & 0.0690 & \textbf{0.0556} \\ \midrule
\multirow{4}{*}{\textbf{1D-CNN}} & Recall & \textbf{0.9811} & 0.9780 & 0.9836 & \textbf{0.9863} & \textbf{0.9424} & 0.8560 & \textbf{0.3333} & 0.1852 & 0.6442 & \textbf{0.6678} & \textbf{0.6962} & 0.6953 \\
& Precision & \textbf{0.9847} & 0.8860 & \textbf{0.9861} & 0.9180 & \textbf{0.9714} & 0.8687 & 0.8438 & \textbf{0.9375} & \textbf{0.7946} & 0.7318 & \textbf{0.7827} & 0.7188 \\
& F1 & \textbf{0.9829} & 0.9297 & \textbf{0.9849} & 0.9509 & \textbf{0.9567} & 0.8623 & \textbf{0.4779} & 0.3093 & \textbf{0.7115} & 0.6983 & \textbf{0.7370} & 0.7069 \\
& FPR & \textbf{0.0029} & 0.0238 & \textbf{0.0054} & 0.0341 & \textbf{0.0048} & 0.0223 & 0.0020 & \textbf{0.0004} & \textbf{0.1678} & 0.2467 & \textbf{0.1389} & 0.1955 \\ \midrule
\multirow{4}{*}{\textbf{Transformer}} & Recall & 0.9931 & \textbf{0.9932} & 0.9928 & \textbf{0.9980} & \textbf{0.9921} & 0.9912 & \textbf{0.6420} & 0.5432 & 0.8480 & \textbf{0.8941} & 0.7903 & \textbf{0.8527} \\
& Precision & \textbf{0.9933} & 0.9727 & \textbf{0.9963} & 0.9860 & \textbf{0.9901} & 0.9845 & 0.7879 & \textbf{0.9565} & 0.8867 & \textbf{0.8905} & \textbf{0.8794} & 0.8653 \\
& F1 & \textbf{0.9932} & 0.9828 & \textbf{0.9946} & 0.9920 & \textbf{0.9911} & 0.9879 & \textbf{0.7075} & 0.6929 & 0.8670 & \textbf{0.8923} & 0.8325 & \textbf{0.8590} \\
& FPR & \textbf{0.0013} & 0.0053 & \textbf{0.0014} & 0.0055 & \textbf{0.0017} & 0.0027 & 0.0056 & \textbf{0.0008} & \textbf{0.1092} & 0.1108 & \textbf{0.0779} & 0.0954 \\ \midrule
\multirow{4}{*}{\textbf{DF}} & Recall & \textbf{0.9969} & 0.9913 & \textbf{0.9977} & 0.9936 & \textbf{0.9950} & 0.9814 & \textbf{0.6914} & 0.5556 & 0.8660 & \textbf{0.8800} & 0.8568 & \textbf{0.8839} \\
& Precision & \textbf{0.9967} & 0.9883 & \textbf{0.9980} & 0.9863 & \textbf{0.9978} & 0.9963 & \textbf{0.9655} & 0.9184 & \textbf{0.8959} & 0.8924 & 0.9135 & \textbf{0.9167} \\
& F1 & \textbf{0.9968} & 0.9898 & \textbf{0.9979} & 0.9899 & \textbf{0.9964} & 0.9888 & \textbf{0.8058} & 0.6923 & 0.8807 & \textbf{0.8862} & 0.8843 & \textbf{0.9000} \\
& FPR & \textbf{0.0006} & 0.0022 & \textbf{0.0008} & 0.0054 & \textbf{0.0004} & 0.0006 & \textbf{0.0008} & 0.0016 & \textbf{0.1015} & 0.1070 & 0.0583 & \textbf{0.0578} \\ \midrule
\multirow{4}{*}{\textbf{BAPM}} & Recall & 0.9937 & \textbf{0.9932} & \textbf{0.9974} & 0.9935 & \textbf{0.9935} & 0.9856 & \textbf{0.7407} & 0.5432 & \textbf{0.8475} & 0.8304 & \textbf{0.8220} & 0.8185 \\
& Precision & \textbf{0.9955} & 0.9301 & \textbf{0.9948} & 0.9249 & \textbf{0.9938} & 0.8637 & \textbf{0.7792} & 0.7458 & 0.8539 & \textbf{0.8711} & 0.8773 & \textbf{0.8799} \\
& F1 & \textbf{0.9946} & 0.9606 & \textbf{0.9961} & 0.9580 & \textbf{0.9937} & 0.9206 & \textbf{0.7595} & 0.6286 & \textbf{0.8507} & 0.8503 & \textbf{0.8487} & 0.8481 \\
& FPR & \textbf{0.0009} & 0.0141 & \textbf{0.0020} & 0.0312 & \textbf{0.0011} & 0.0268 & 0.0068 & \textbf{0.0060} & 0.1462 & \textbf{0.1238} & 0.0826 & \textbf{0.0803} \\ \midrule
\multirow{4}{*}{\textbf{TMWF}} & Recall & \textbf{0.9980} & 0.9937 & \textbf{0.9974} & 0.9933 & \textbf{0.9942} & 0.9670 & \textbf{0.7901} & 0.5185 & \textbf{0.8918} & 0.8798 & 0.8645 & \textbf{0.8757} \\
& Precision & \textbf{0.9944} & 0.9625 & \textbf{0.9990} & 0.9951 & \textbf{0.9973} & 0.9902 & \textbf{0.8205} & 0.7778 & 0.8935 & \textbf{0.9022} & 0.9080 & \textbf{0.9124} \\
& F1 & \textbf{0.9962} & 0.9779 & \textbf{0.9982} & 0.9942 & \textbf{0.9957} & 0.9785 & \textbf{0.8050} & 0.6222 & \textbf{0.8926} & 0.8909 & 0.8857 & \textbf{0.8937} \\
& FPR & \textbf{0.0011} & 0.0073 & \textbf{0.0004} & 0.0019 & \textbf{0.0005} & 0.0016 & 0.0056 & \textbf{0.0048} & 0.1072 & \textbf{0.0961} & 0.0630 & \textbf{0.0604} \\ \midrule

\multirow{4}{*}{\textbf{LUCID}} 
& Recall    & \textbf{0.8421} & 0.8360 & \textbf{0.9641} & 0.8651 & \textbf{0.9420} & 0.9404 & 0 & 0 & 0 & 0 & \textbf{0.3647} & 0.3529 \\
& Precision & 0.9372 & \textbf{0.9454} & 0.9296 & \textbf{0.9481} & 0.9680 & \textbf{0.9825} & 0 & 0 & 0 & 0 & \textbf{0.7029} & 0.6957 \\
& F1        & 0.8871 & \textbf{0.8873} & \textbf{0.9465} & 0.9047 & 0.9548 & \textbf{0.9610} & 0 & 0 & 0 & 0 & \textbf{0.4803} & 0.4683 \\
& FPR       & 0.0048 & \textbf{0.0041} & 0.0150 & \textbf{0.0097} & 0.0123 & \textbf{0.0066} & -- & -- & -- & -- & \textbf{0.0336} & 0.0337 \\ \midrule

\multirow{4}{*}{\textbf{Whisper}} 
& Recall    & \textbf{0.9879} & 0.9800 & \textbf{0.9933} & 0.9821 & 0.9922 & \textbf{0.9927} & 0 & 0 & \textbf{0.8572} & 0.8411 & 0.0888 & \textbf{0.6236} \\
& Precision & \textbf{0.9924} & 0.9482 & \textbf{0.9914} & 0.9612 & \textbf{0.9955} & 0.9690 & 0 & 0 & 0.8368 & \textbf{0.8507} & 0.5300 & \textbf{0.9006} \\
& F1        & \textbf{0.9902} & 0.9638 & \textbf{0.9924} & 0.9715 & \textbf{0.9939} & 0.9807 & 0 & 0 & \textbf{0.8469} & 0.8459 & 0.1521 & \textbf{0.7369} \\
& FPR       & \textbf{0.0006} & 0.0046 & \textbf{0.0018} & 0.0081 & \textbf{0.0019} & 0.0133 & -- & -- & 0.1794 & \textbf{0.1584} & 0.0360 & \textbf{0.0315} \\ \midrule

\multirow{4}{*}{\textbf{MATEC}} 
& Recall    & 0.9905 & \textbf{0.9951} & 0.9932 & \textbf{0.9962} & 0.9946 & \textbf{0.9963} & 0.6790 & \textbf{0.7654} & \textbf{0.8931} & 0.8923 & 0.8481 & \textbf{0.8739} \\
& Precision & 0.9961 & \textbf{0.9983} & 0.9972 & \textbf{0.9980} & 0.9984 & \textbf{0.9989} & 0.8871 & \textbf{0.8986} & 0.8792 & \textbf{0.9004} & 0.8653 & \textbf{0.8825} \\
& F1        & 0.9933 & \textbf{0.9967} & 0.9952 & \textbf{0.9971} & 0.9965 & \textbf{0.9976} & 0.7692 & \textbf{0.8267} & 0.8861 & \textbf{0.8963} & 0.8566 & \textbf{0.8782} \\
& FPR       & 0.0003 & \textbf{0.0001} & 0.0006 & \textbf{0.0004} & 0.0006 & \textbf{0.0005} & \textbf{0.0028} & \textbf{0.0028} & 0.1237 & \textbf{0.0995} & 0.0949 & \textbf{0.0836} \\ \midrule

\multirow{4}{*}{\textbf{SmartDetector}} 
& Recall    & 0.9989 & \textbf{0.9999} & 0.9999 & \textbf{1} & \textbf{1} & \textbf{1} & 0.6667 & \textbf{0.7160} & 0.8423 & \textbf{0.8906} & 0.8717 & \textbf{0.9033} \\
& Precision & 0.9990 & \textbf{0.9999} & \textbf{0.9999} & 0.9997 & 0.9997 & \textbf{0.9999} & 0.8308 & \textbf{0.9355} & 0.8763 & \textbf{0.8795} & 0.9254 & \textbf{0.9344} \\
& F1        & 0.9990 & \textbf{0.9999} & \textbf{0.9999} & \textbf{0.9999} & 0.9998 & \textbf{0.9999} & 0.7397 & \textbf{0.8112} & 0.8589 & \textbf{0.8850} & 0.8978 & \textbf{0.9186} \\
& FPR       & 0.0001 & \textbf{0} & \textbf{0} & 0.0001 & 0.0001 & \textbf{0} & 0.0044 & \textbf{0.0016} & \textbf{0.1276} & 0.1309 & 0.0321 & \textbf{0.0290} \\

\bottomrule
\end{tabular}
}
\begin{tablenotes}
\setlength{\leftskip}{-1em}
\setlength{\rightskip}{1em}
\small
\item * \textbf{NoAug}: No augmentation. 
\item ** The best results are marked in \textbf{bold}. We mark the FPR as ``--'' when the F1 score is 0 (as FPR is meaningless in this case). 
\end{tablenotes}
\end{table*}

For the window length, a smaller window improves the time-domain resolution but limits the range of frequency components available for mixing. Conversely, a larger window enables more diverse mixing patterns but risks losing some local features. As illustrated in Figure \ref{fig:sensitivity analysis}, the optimal window length varies across models; however, for most models, a window length of 100 leads to relatively lower F1 scores and higher FPRs, highlighting the importance of windowing. For DF, the F1 score does not peak until the window length reaches 50. In contrast, 1D-CNN and LUCID experience significant performance degradation once the window length exceeds 10. Interestingly, for 1D-CNN and LUCID, a noticeable rebound in F1-score occurs starting from a window length of 20. This phenomenon may be attributed to the diversity gains from richer mixing patterns beginning to outweigh the performance degradation caused by reduced time-frequency resolution again. For the majority of models, the FPR tends to rise with increasing window length. This likely occurs because larger windows result in coarser granularity of the mixing perturbations, thereby interfering with the model’s ability to precisely distinguish malicious samples. A window length of 10 provides a reasonable trade-off between feature granularity and mixing diversity, and is therefore recommended in our method.

\end{document}